\documentclass[fleqn,usenatbib]{mnras}
\usepackage{multirow}
\usepackage{rotating}
\usepackage{colortbl}
\usepackage{color}
\usepackage{amsmath}
\usepackage{amsfonts}
\usepackage{verbatim}
\usepackage{scalefnt}
\usepackage[percent]{overpic}
\usepackage[T1]{fontenc} 
\usepackage{aecompl} 
\usepackage{ulem}
\usepackage{cleveref}
\usepackage{numprint}

\usepackage{float}

\def\lsim{\mathrel{\rlap{\lower 3pt \hbox{$\sim$}} \raise 2.0pt \hbox{$<$}}}
\def\gsim{\mathrel{\rlap{\lower 3pt \hbox{$\sim$}} \raise 2.0pt \hbox{$>$}}}

\newcommand{\comments}[1]{} 

\title[Relativistic dynamics and OJ-287's flares]{Relativistic binary-disc dynamics and the timing of OJ-287's flares.}
\author[L. Zwick et al.]{Lorenz Zwick and Lucio Mayer.
\\
Center for Theoretical Astrophysics and Cosmology, Institute for Computational Science, University of Zurich,\\ Winterthurerstrasse 190, CH-8057 Z{\"u}rich, Switzerland}

\date{Accepted XXX. Received YYY; in original form ZZZ}

\pubyear{2022}

\begin{document}

\pagerange{\pageref{firstpage}--\pageref{lastpage}}

\maketitle


\begin{abstract}
We revisit the precessing black hole binary model, a candidate to explain the bizarre quasi-periodic optical flares in OJ-287's light curve, from first principles. We deviate from existing work in three significant ways: 1) Including crucial aspects of relativistic dynamics related to the accretion disc's gravitational moments. 2) Adopting a model-agnostic prescription for the disc's density and scale height. 3) Using monte-carlo Markhov-chain methods to recover reliable system parameters and uncertainties. We showcase our model's predictive power by timing the 2019 Eddington flare within 40 hr of the observed epoch, exclusively using data available prior to it. Additionally, we obtain a novel direct measurement of OJ-287's disc mass and quadrupole moment exclusively from the optical flare timings. Our improved methodology can uncover previously unstated correlations in the parameter posteriors and patterns in the flare timing uncertainties. According to the model, the 26th optical flare is expected to occur on the 21st of August 2023 $\pm$ 32 days, shifted by approximately a year with respect to previous expectations.
\end{abstract}

\begin{keywords}
quasars: supermassive black holes -- black hole physics -- accretion:accretion discs -- methods: data analysis.
\end{keywords}


\section{Introduction}
\label{sec:intro}
The precessing black hole binary model (hereafter PBM\footnote{In this paper, the acronym PBM specifically refers to the model developed and refined by M.Valtonen and several collaborators over the last two decades. It does not refer to the general idea of a precessing binary causing features in OJ 287's light curve.}) originally presented in \citet{LV96} has arguably been the most successful in explaining and predicting several unique features of OJ-287's luminosity curve \citep{1916Wölf,1971Browne,1971Kinman,1973Craine,1984Corso}. With the first observations dating to the late 1880ies, this peculiar object is now classified as a blazar, situated at a redshift of $z=0.306$ \citep{1985sitko,2003Carangelo}, and is thus composed by a supermassive black hole surrounded by an accretion disc powering a relativistic jet \citep{1993antonucci,1995Urry,1998Ghisellini,2003dunlop}. Crucially, OJ 287's light curve features bright, doubly peaked optical flares occurring with an approximate periodicity of twice every 12 years, as well as a slower modulation on a timescale of $\sim 60$ yr \citep{2000Valtaoja,2010fan,2014JTang}. Inspired by the original work by \cite{1988Sillanpää}, the PBM proposes a scenario in which the periodicity is explained by the presence of a smaller secondary black hole, orbiting the primary on a highly relativistic, inclined and eccentric trajectory that can be matched to the available data \citep[see also][]{1994karas}. The sharp optical flares are then associated to impacts between the secondary and the disc, producing the characteristic twice per 12 yr orbital period structure. The 60 yr modulation is instead associated to the relativistic periastron advance timescale of the binary, thus fixing the system's characteristic mass and size to $\sim 10^{10}$ M$_{\odot}$ and $\sim 0.1$ pc, respectively.

From its original inception, the PBM has since gone through several iterations and improvements, including for example a more sophisticated description of the disc's response to the impacts \citep{2006valtonen2,2006valtonen}, the use of more detailed post-Newtonian equations of motion \citep{2010valtonen,2016valtonen,2022kacskovics}, as well as incorporating input from numerical simulations \citep{1997Sundelius} and ulterior electromagnetic data \citep{1997Yanny,2018Dey,2023titarchuk}. The strength of this model is exemplified in the confirmation by the Spitzer Space Telescope of the so called "Eddington flare" \citep{2020Laine}, within a truly remarkable 4 hr of the predicted epoch: the 31st of July 2019 UT $\pm$ 4.4 hr \citep{2018Dey}. {More recently the model has been challenged by the works of \citep{2023komossa,2023komossa2}, based on both the alleged lack of a flare in October 2022 and the use luminosity scaling relations. This specific date of the 2022 flare is mentioned in the pre-print ArXiv document \citet{2022valtonen}, while the different prediction of July is given in both previous \citep{2007valtonen} and later \citep{2023valtonen} published works, based on a more sophisticated treatment of the accretion disc's scale height.}


Notwithstanding the claims in \cite{2023valtonendiss}, the recent debate and the inherent complexity of OJ-287's system \citep[along with some interesting discrepancies also discussed in e.g.][]{2023komossa,2023komossa2} warrant to take a second look at alternative models for OJ 287's light curve. Many of the ones proposed in the literature still invoke the presence of a secondary black hole \citep{1997katz, 1998Villata,2000Valtaoja}, but do not require such highly relativistic initial conditions. Other forgo the requirement of a binary entirely, explaining the variability via complex jet beaming and precession effects \citep{2010Villforth,2018qian,2018Britzen,2020butozova}. All of the alternative models above are physically plausible and able to qualitatively explain many features of the blazar's emission, including variability in bands other than optical \citep[see][in particular]{2000Valtaoja}. However, it is important to note that, as of today, only the PBM has been able to repeatedly predict the timing of optical flares \citep[occasionally with spectacular precision, see e.g.][]{2017gupta,2018Dey,2020Laine}, and that some of the many additional components of OJ-287's light curve \citep[see e.g.][]{1985valtaoja,1990dediego,2013pihajoki,2013pihajoki2} can still be associated to a highly precessing binary.

{In this work,} we set out to revisit the PBM from first principles, deviating from existing methodology in three significant ways: Firstly, we model the dynamical effects of the accretion disc's gravitational moments and their post-Newtonian cross terms (see Section \ref{sec:sub:eom}), a crucial element that was entirely missing in previous PBM iterations. Secondly, we adopt an agnostic description of OJ-287's accretion disc based on density and scale-height power laws, as opposed to assuming any specific model. Finally, we use Bayesian monte-carlo Markhov-chain methods (hereafter MCMC, see Section \ref{sec:sub:fit}) to recover reliable posterior distributions for the system's parameters. As a consequence of our revised methodology, we obtain significantly diverging results from the established literature, which are discussed throughout Section \ref{sec:results}.

\begin{figure}
    \centering
    \includegraphics[scale=0.28]{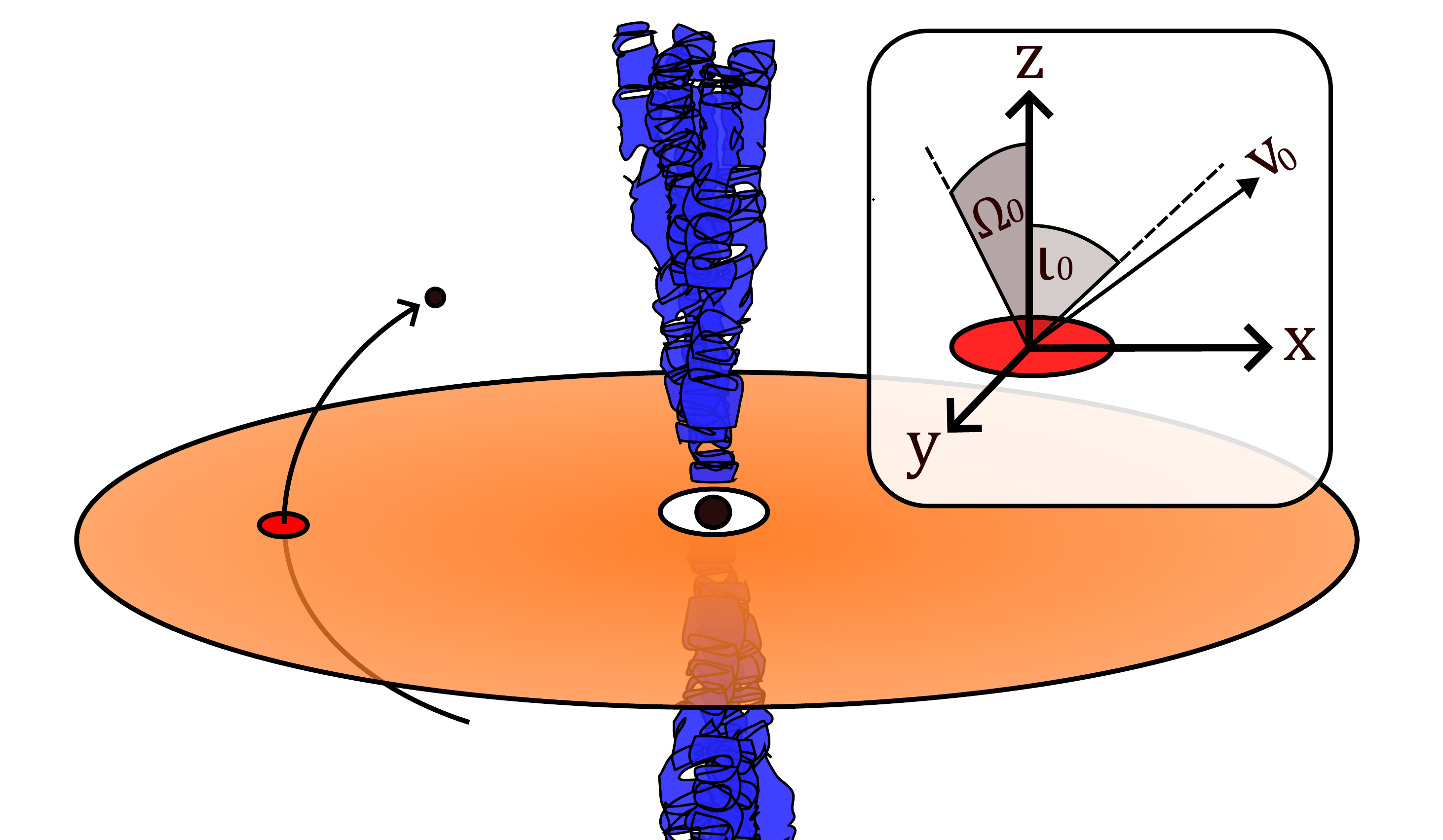} 
    \caption{A simple cartoon of the OJ-297 system for visualisation purposes. In our integrations, we initialise the secondary black hole at the position (-$x_0,0,0$), with a velocity vector specified by it magnitude $v_0$, its inclination with respect to the z-x plane $\Omega_0$ and its tilt towards the central black hole $\iota_0$.}
    \label{fig:sketch}
\end{figure}

\section{Methodology}
\label{sec:methods}
\subsection{Basic setup and historical flare timings}
\label{sec:sub:basic}
In both the PBM and our work, the basic setup of OJ-287's system consists in a primary black hole of mass $M$ surrounded by a thin accretion disc, as illustrated in Figure \ref{fig:sketch}. A smaller secondary black hole of mass $m$ orbits the primary on a highly eccentric, inclined orbit that intersects the disc twice every orbital period. The impact between the secondary and the disc deposits a large amount of energy into a localised region of gas \citep[by means of relativistic Bondi accretion,][]{1952Bondi,2011Zanotti}, which subsequently releases a flare of electromagnetic radiation. Given this basic picture, there are currently 11 detected optical flares in OJ 287's light curve that can be clearly associated with impacts between the secondary and the disc \citep{2021Dey,2022valtonen,2023valtonen}. Their epochs are listed together with the corresponding uncertainties in Table \ref{tab:flares}.

Over many refinements, the PBM has attempted to fit and predict the timings of the impact flares with higher and higher precision, eventually resulting in some extremely tight constraints for many parameters of the OJ-287 system, which we report in Table \ref{tab:parameters}). Such tight bounds are made possible by the precision with which the historical flare timings have been observed (see Table \ref{tab:flares}), which occasionally amounts to a timing uncertainty below a few hours, or more typically $\sim 0.01$ yr. In general, the observations vary in their precision as a consequence of the available instruments and the broadness of the flare's luminosity curves, among other experimental constraints \citep{2020Laine,2021Valtonen}. \newline \newline
The goal of this work is to re-derive a precessing binary model for OJ-287's system while also addressing some possible shortcomings of the PBM. The first ingredient is an extremely accurate description of the binary dynamics, collecting all contributions to the equations of motion that cause timing shifts comparable to the reported uncertainties (see Section \ref{sec:sub:eom}). Then, we have to account for possible astrophysical mechanisms that can delay the emission of radiation after the impact occurs (see Section \ref{sec:sub:delays}). Finally, we have to devise a strategy to efficiently explore a large parameter space of initial conditions, in order to fit the historical flare timings and recover parameters reliably (see Section \ref{sec:sub:fit}). For the purposes of this work, we settle on a target accuracy of $\sim 0.01$ yr {regarding the dynamical evolution of the system}, over a total integration time of approximately one century, close to the typical uncertainties in the data. Furthermore, we will show how the PBM is lacking of a proper description of the accretion disc's gravitational potential, a modelling oversight that can induce timing shifts of order months to years, over the total integration time of approximately 1 century. A short list summarising the differences between our model and the PBM cand be found in Table \ref{tab:model}.

\begin{table}
    \centering
    \begin{tabular}{c|c|c}
      Nr. & Epoch  & Uncertainty [yr]  \\
      \hline
        6 & 1912.980 & $\pm$0.02 \\
        12 & 1947.283 & $\pm$0.002 \\
        13 & 1957.095 & $\pm$0.025 \\
        17 & 1972.935 & $\pm$0.012 \\
        18 & 1982.964 & $\pm$0.0005 \\
        19 & 1984.125 & $\pm$0.01 \\
        21 & 1995.841 & $\pm$0.002 \\
        22 & 2005.745 & $\pm$0.015 \\
        23 & 2007.692 & $\pm$0.0015 \\
        24 & 2015.875 & $\pm$0.025 \\
        25 & 2019.569 & $\pm$0.0005 
    \end{tabular}
    
    \caption{Epochs and uncertainties of the 11 detected impact flares in the light curve of the blazar OJ 287, as compiled by \citet{2018Dey} and \citet{2020Laine}. The numbering begins by convention at the value 6, as evidence for flares earlier than 1912 has been found in archival photographic plates dating to 1886 \citep{1982gaida,2013Hudec}. Large data gaps are caused by the events of World War I and II, while later occasional gaps are due to observability constraints at solar conjuction.}
    
    \label{tab:flares}
\end{table}

\subsection{Binary, disc and the equations of motion}
\label{sec:sub:eom}
\subsubsection{Post-Newtonian Binary Dynamics}
Taking the values reported in \citet{2018Dey} as a baseline, the black hole binary in OJ-287 can be characterised by a mass $M \sim 2 \times 10^{10}$ M$_{\odot}$, a secondary to primary mass ratio $q = m/M \sim 0.01$ and a semi-latus rectum $p \sim 20 \, r_{\rm S}$, where we defined the Schwarzschild radius of the primary $r_{\rm S} = 2 G M c^{-2}$. In order to precisely match the observed flare timings, we are required to model the dynamics of this highly relativistic binary with sufficient precision. As already noted in many PBM papers, a possibility is to use post-Newtonian (hereafter PN) equations of motion (hereafter EoM). We can estimate the required PN order by the following simple considerations: corrections to Newtonian binary dynamics scale phenomenologically as powers of the dimensionless quantity $r_{\rm{S}}/p$, i.e. the system's Schwarzschild radius divided by the typical time-averaged orbital size \citep{1938einstein,2014blanchet,2017iorio,2018maggiore,2018schaefer,2021zwick}. Over an integration time of $\sim 100$ yr, it is therefore expected that PN corrections would produce a total accumulated shift in the flare timing of roughly $100/20^n $ yr, where $n$ is the order of the correction. Therefore, third order PN corrections are required to achieve a timing accuracy of $0.01$ yr. {Beyond this simple estimate, we explicitly show that the higher order PN terms used in \citet{2018Dey} and \citet{2020Laine} are subdominant with respect to the modifications induced by the disc potential in Fig.~\ref{fig:accelerations}.} For the purposes of our work, we adopt the standard PN EoM for isolated binary black holes, which take a convenient form when expressed as an acceleration \citep[see e.g.][for the explicit coefficients]{2014blanchet,2017will,2018bernard}:
\begin{align}
    \frac{d \mathbf{v}}{dt} = -\frac{G M_{\rm{tot}}}{r^2}\left[\left( 1 + \mathcal{A} \right) \mathbf{n}  + \mathcal{B} \mathbf{v}\right],
\end{align}
where $\mathbf{v}$ is the reduced mass' velocity vector, $\mathbf{n}$ its unit vector while the coefficients $\mathcal{A}$ and $\mathcal{B}$ contain various PN modifications to the inverse square law. Of particular note are the 1.5 PN corrections required to model the primary's spin \citep{1975barker,1981barker}, and the 2.5 PN corrections enforcing the the slow decay of orbital parameters caused by radiation reaction \citep{1963peters}. {The precession of the primary's spin vector $\Vec{S}_1$ is mediated by the secondary's mass \citep{1981barker,1995kidder,2006porto,2006faye}. At lowest order in PN theory it scales dimensionally as:
\begin{align}
    \dot{\Vec{S}}_1 \sim \frac{G m v }{c^2 r^2}.
\end{align}
It is therefore suppressed by a factor $\sim 100$ with respect to other first order PN terms. The timescale on which the primary's spin evolves is thus approximately 100 times longer than the relativistic perihelion precession timescale, which already amounts to $\sim 120$ yr for the system parameters. Over the available observation time, the primary's spin vector may thus at most precess by a few degrees. We show explicitly in Fig.~\ref{fig:accelerations} that the contributions to the timings caused by such small changes in spin orientation are subdominant with respect to the effect of the disc's potential, and we therefore neglect them. Note that \citet{2018Dey} takes a more sophisticated approach, modelling the orientation and the evolution of the central BH's spin vector with PN equations. However, also note that \cite{2018Dey} may be overestimating the magnitude of high order PN terms, in particular radiation reaction. In the latter work it is stated that the 2.5 PN terms cause a change in the orbital period of approximately $\sim 10^{-3}$ per orbit. Taking the notorious evolution equation from \citet{1963peters}:
\begin{align}
    \label{eq:peters}
    \frac{\dot{a}}{a} = -\frac{64}{5}\frac{(G M)^3}{c^5a^4}\frac{q}{(1+q)^2} f(e),
\end{align}
where $q$ is the mass ratio and $f(e)$ the eccentricity enhancement function. Evaluating the equations for a mass of 1.8$\times10^{10}$ M$_{\rm{\odot}}$, a mass ratio of $q\sim100$, a semimajor axis of $\sim 50 \, r_{\rm S}$ and an eccentricity of $\sim 0.7$, we find a period loss per orbit of $\sim 5\times 10^{-5}$, which is also is also consistent with the numerical integrations shown in Fig.~\ref{fig:accelerations}\footnote{It is possible that Eq.~\ref{eq:peters} may have been evaluated for equal mass binaries in \cite{2018Dey}.}}. Radiation reaction terms of even higher order will only weakly affect the dynamics of the system, as their strength is similarly suppressed by the small mass ratio. According to the simple estimate detailed above, the 3.5 PN radiation reaction term would only contribute to a twenty minute shift over an integration time of 100 yr, which is actually comparable to the reported $\sim$ hr shift in \citet{2020Laine}. 
Schematically, our binary EoM can be summarised as follows:
\begin{align}
    \frac{d \mathbf{v}}{dt} &= \text{N} + \text{1PN} +  \text{2PN}  + \text{3PN} &\text{ Binary} \\ 
    &+ \text{1.5PN} &\text{ Spin-orbit} \\
    &+ \text{2.5PN}& \text{ Radiation reaction.}
\end{align}
To fully specify these equations in an appropriate frame of reference we have to provide three parameters, i.e. the primary's mass $M$, the secondary's mass $m$ and the primary's dimensionless spin parameter $\xi$.

\subsubsection{Disc Multipoles and Cross Terms}
Newtonian and PN point mass forces are not the only relevant contributions to this system's dynamics. According to the estimates in \citet{2019valtonendisc}, derived self-consistently in the PBM, the accretion disc surrounding OJ-287's primary has a typical scale height of $\sim 10^{15}$ cm and a typical number density of $\sim 10^{14}$ cm$^{-3}$. Taking the values mentioned above, we obtain a reference gas mass enclosed within $\sim 100 \, r_{\rm{S}}$ of order several $10^{8}$ M$_{\odot}$, close to $1 \%$ of the system's total mass. Clearly, the gravitational influence of such a massive disc is an important ingredient to faithfully capture the dynamics of the binary, as it can potentially produce timing shifts of order 1 yr over the expected integration time.
For the purposes of this work, we adopt a simple parametrised disc model characterised by a density profile $\rho$ and a scale height profile $h$. Crucially, we do not assume that the disc structure is determined by the $\alpha$-viscosity prescription \citep{1973shakura}, as is customary in the PBM. Rather, we allow for the two profiles to be independent power laws:
\begin{align}
    \rho(R) &= \rho(l_{\rm s}) \left(\frac{l_{\rm s}}{R}\right)^{j_1}\\
    h(R) &= h(l_{\rm s}) \left(\frac{l_{\rm s}}{R}\right)^{j_2},
\end{align}
where $l_{\rm{s}}$ is an arbitrary length scale (for example the innermost stable circular orbit at $3r_{\rm{S}}$) and $R$ is the radial distance to the primary. We model the disc's potential as a mean field gravitational monopole (DM) and a mean field gravitational quadrupole (DQ). Toghether, the two capture the fundamental influence of the disc on the trajectory of the binary, i.e. an additional radial force as well as an axisymmetric perturbation. Additionally, we make the assumption that the primary's spin vector is closely aligned with the disc's symmetry axis, as expected for a black hole that has grown through gas accretion \citep{1998natarajan,2005volonteri,2005king,2012barausse}.

Decomposing the disc's potential allows us to write down explicit analytical formulae for the resulting accelerations, rather than having to perform numerical integrations of the disc's mass distribution. Crucially, the latter would dramatically slow down likelihood function evaluations in our numerical pipeline (see section \ref{sec:sub:fit}). Another option would be to use analytical disc potentials. However, only few density-potential pairs exist \citep{1964toomre,1987Kuzmin,1987binney}, defeating the purpose of being as general as possible. Thus, we adopt EoM of the following form:
\begin{align}
    \frac{d \mathbf{v}}{dt}\lvert_{\rm{DM}} &= - \frac{GM_{\rm{d}}}{r^2}\mathbf{n} \\
     \frac{d \mathbf{v}}{dt}\lvert_{\rm{DQ}} &= - \frac{3 G Q_2}{2 r^4} \left[5\mathbf{n}(\mathbf{e}_{\rm z} \cdot \mathbf{n}) - 2 \mathbf{e}_{\rm z}(\mathbf{e}_{\rm z} \cdot \mathbf{n}) - \mathbf{n}  \right],
\end{align}
where we aligned they system's symmetry axis with the z-direction, and $\mathbf{e}_{\rm z}$ is a unit vector. Note that, for similar reasons as originally presented in \citet{LV96}, we can neglect subdominant frictional and accretion forces acting on the secondary while it is submerged within the disc. The disc's enclosed mass $M_{\rm d}$ and mass quadrupole $Q_2$ are defined as:
\begin{align}
   M_{\rm d}(r) &= 2 \pi\int^{r}_{0}\int^{h(r')}_{-h(r')} r' \rho(r') \, dz' dr' \\
   Q_{2}(r) &= 2 \pi\int^{r}_{0}\int^{h(r')}_{-h(r')} r' \rho(r')\left(2z'^2 - r'^2 \right) \, dz' dr'
\end{align}
Both of these quantities only depend on three different combinations of the density and scale height profile parameters. For convenience, we thus define an enclosed mass profile and a dimensionless quadrupole moment $J_2$:
\begin{align}
    M_{\rm d}(r) &= M_{\rm d}(l_{\rm{s}}) \frac{r^2}{l_{\rm{s}}^2}\left(\frac{l_{\rm s}}{r} \right)^{j_{\rm{eff}}} \\
    J_2 &= \frac{Q_2(r)}{ M_{\rm d}(r)r^2}
\end{align}
where:
\begin{align}
    M_{\rm d}(l_{\rm{s}}) & = \frac{4 \pi}{2-j_{\rm eff}} h(l_{\rm{s}}) \rho(l_{\rm s})l_{\rm s}^2,
\end{align}
and $j_{\rm{eff}} = j_1 + j_2$. Then, the three parameters $M_{\rm{d}}$, $j_{\rm{eff}}$ and $J_2$ completely specify the gravitational mean field monopole and quadrupole of the disc. Note that in general, the disc's monopole contribution is \textit{not} degenerate with the primary's mass, because its influence on an eccentric test particle varies with the orbital phase. Note also that, in the limit of thin discs, the dimensionless quadrupole moment $J_2$ reduces to recognisable formula:
\begin{align}
    \label{eq:jeff}
    J_2 = -\frac{j_{\rm{eff}} -2}{j_{\rm{eff}} -4},
\end{align}
leading to the classic result of $J_2 = -1/2$ for a flat, thin homogeneous disc.

Our model for OJ-287's accretion disc comes at a cost of one extra free parameter with respect to the the PBM, in which the disc is specified by a choice of the viscosity $\alpha$ and accretion rate $\dot{M}$. While the latter choice certainly follows astrophysical expectations \citep[at least as an effective model, see e.g.][often with values of $\alpha \sim 0.1$]{2004gierli,2007king,2012kotko}, relying on the $\alpha$-disc model requires solving structural and thermal equations to obtain density and scale height profiles. In the context of the PBM, these equations have typically been separately solved beforehand, invoking a specific numerical model based on the work of \citet{1981sakimoto} and \citet{1984rosner}. Considering how widely the behaviour of simulated accretion discs changes as a function of the chosen opacity tables or viscosity prescription \citep{2013abramowicz,2016Jiang,2020jiang,2023chen}, trading the computational and theoretical overhead of the PBM for a more general approach is of very clear benefit, despite the additional parameter. {The downside of our choice is the impossibility to analytically model the rich dynamical interactions of the disc with the secondary, which include disc warping and scale height changes. The latter effects have been observed in hydrodynamical simulations of the system, and are required for the PBM to successfully match the flare timing in OJ-287's light curve \citep{2023valtonen,2023valtonendiss}. To our own surprise, we find that our model is still able to reproduce the correct timings optical flares without requiring a time advance prescription.}  \newline\newline
Finally, a consistent description of a highly relativistic PN system with monopole and quadrupole perturbations requires the inclusion of several additional cross terms. The importance of PN-cross terms has been thoroughly discussed in \citet{2014will2} and \citet{2014will}. In short, they are fundamental to assure that the system's energy is properly conserved over relativistic perihelion advance timescales, a factor that is crucial when one hopes to track the timing of consecutive flares. Thus, we must add both monopole and quadrupole PN cross terms in our EoM, up to first PN order \citep[see][for the explicit formulae]{ 2014will2,2014will}.

\subsubsection{EoM integration and plane intersections}
\begin{figure}
    \centering
    \includegraphics[scale=0.70]{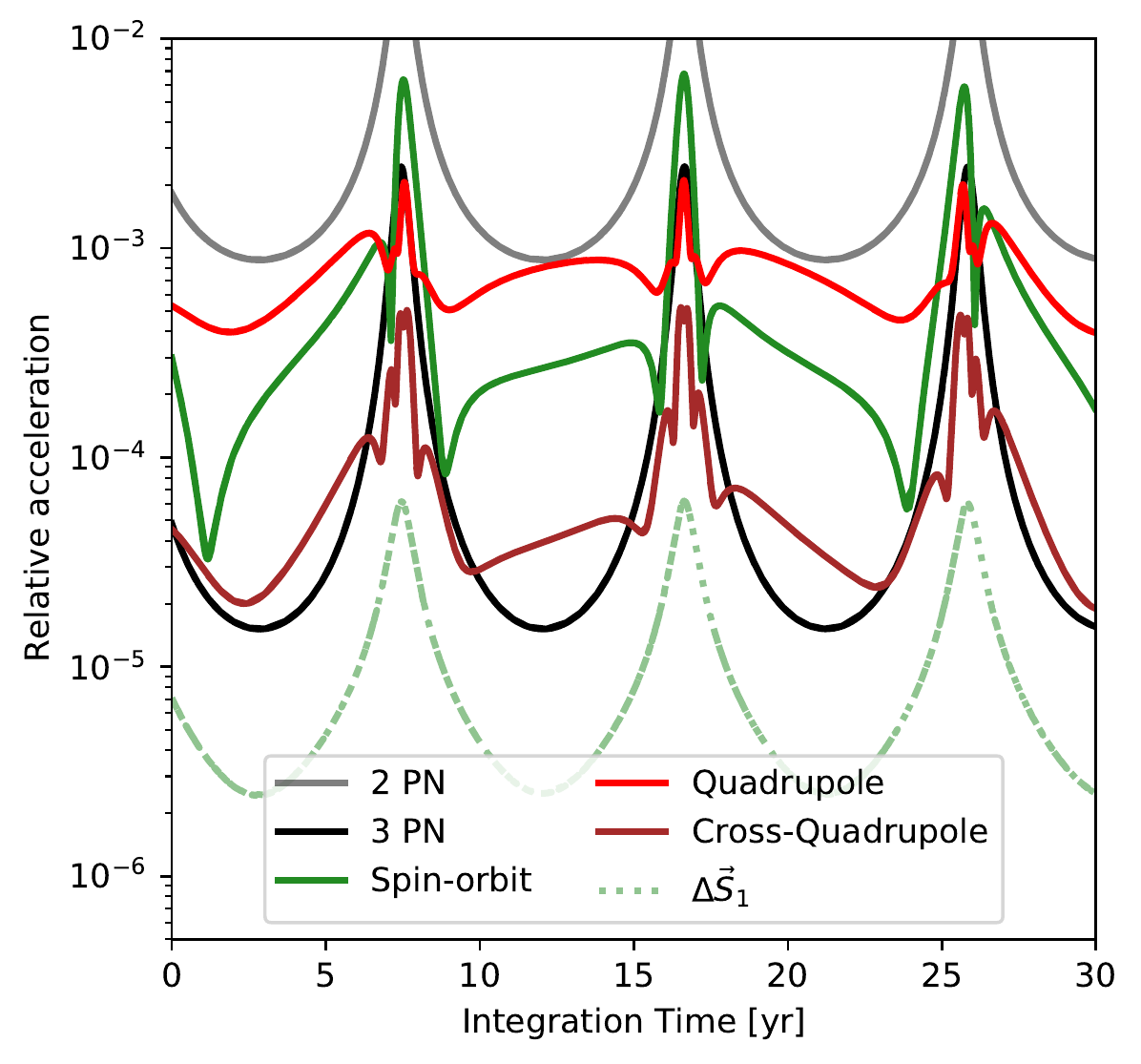} 
    \caption{Magnitude of several different components of the conservative EoM relative to the Newtonian acceleration. The system is integrated for a few orbits given the initial conditions and best fit parameters reported in Table \ref{tab:posteriors}. Note how the accretion disc's quadrupole potential and its PN cross term \citep[red  and brown lines, see][]{2014will2,2014will} cause gravitational forces that are {comparable or higher than than high order PN forces. Yet, they have been neglected when modelling OJ-287 before this work. The dotted pale green line shows the difference in acceleration caused varying the primary spin direction by 1 degree, clearly showing how the effects of the latter's precession are negligible over the observation time.} }
    \label{fig:accelerations}
\end{figure}

To summarise, we adopt a set of EoM of the following schematic form:
\begin{align}
    \frac{d \mathbf{v}}{dt} &= \text{N} + \text{1PN} + \text{1.5PN} + \text{2PN}  + \text{3PN} &\text{ Conservative} \\
    &+ \text{2.5PN}& \text{ Dissipative}\\ &+ \text{DM} +\text{DQ}  & \text{ Disc}  \\ &+ \text{DM}\times\text{1PN} + \text{DQ}\times\text{1PN} & \text{ Cross terms.}
\end{align}
Relative to the Newtonian contribution, we expect conservative PN accelerations to be smaller by a factor of roughly $\sim 20^n$ on average over an orbit, where $n$ is the PN order. The 2.5PN contribution is further suppressed by a factor $\sim100$ due to the mass ratio, making it de-facto the least important term in the adopted EoM. In addition to point mass terms, the disc monopole and quadrupole both can contribute at $\sim 1\%$ relative level, given the baseline values in \citet{2019valtonendisc}. The PN cross terms are suppressed by a further factor $\sim 20$, contributing at the $\sim 0.1\%$ level. Note that these rough expectations are confirmed a posteriori throughout our numerical integrations, an example of which is shown in Figure \ref{fig:accelerations}. Crucially, both disc and cross term contributions to the EoM had previously been neglected in the PBM, depite being orders of magnitude larger than even the 2.5PN radiation reaction force. \newline \newline
Given the EoM and a set of initial conditions, our first goal is to efficiently determine the time at which the secondary black hole impacts the disc, or equivalently intersects the $z = 0$ plane. To achieve this, we integrate the EoM with the 8th order Runge-Kutta implementation "DOP853" of the scipy \texttt{solve\_ivp} package \citep{2020scipy} and extract the $z$ component of the secondary's position vector. To find the epochs of the impacts, we search for all local maxima of the function $\lvert 1/z \rvert$ with the scipy \texttt{find\_peaks} function, further refining the impact times with the in-built interpolation feature "dense\_output" of \texttt{solve\_ivp}. We perform several resolution studies, finding that a choice of the integrator parameter $r_{\rm{tol}} = 10^{-5}$ and a temporal interpolation resolution of 120 hr assures that the numerical error in the impact timings remains below $\sim 1$ hr for all reasonable initial conditions. A single run of the integration and interpolation pipeline requires approximately 0.8 s of computation time, as opposed to several minutes when the same accuracy is requested without interpolation.

\subsection{Time Delays}
\label{sec:sub:delays}
\subsubsection{Disc Delay}
As a result of the impact of the secondary with the disc, a hot bubble of gas forms, expands and releases a flare of thermal radiation after a certain time delay.
For the purposes of this work, we settle on the original delay prescription used in \citet{LV96}, derived from theoretical principles tailored to the radiation dominated inner regions of accretion discs. As shown more recently, the physics upon which the prescription is based are also able to reproduce the luminosity, spectrum and duration of the optical impact flares \citep{2016pihajoki,2019dyba}, and may also be a plausible explanation for quasi periodic emissions \citep{2023franchini}. Note however that several different prescriptions for such delays have been experimented with throughout the early iterations of the PBM \citep{LV96,1998ivanov}. A thorough and interesting discussion on the effects of varying the delay prescription can be found in \citet{1998Pietila}. The disc time delay, $\tau_{\rm d}$, depends on both the properties of the disc at the impact site, as well as the properties of the impactor:
\begin{align}
    \label{eq:delay}
    \tau_{ \rm d} \propto h^{13/21} m^{22/21} \rho^{51/56} r^{355/168} \delta^{-355/84},
\end{align}
where the parameter $\delta$ describes the impactor's orientation and velocity relative to a Keplerian disc \citep{2016pihajoki}. Similarly to \citet{2018Dey}\footnote{In \citet{2018Dey}, this additional free parameter was associated to the disc's scale height rather directly with the time delay.}, we preserve the physical scaling laws of Eq. \ref{eq:delay}, but allow it to be modified by a constant proportionality pre-factor $f_{\rm{err}}$, which is fitted as part of the orbital solution. Allowing for this freedom is important, since the disc's response to impacts is the most complex hydro and thermo-dynamical ingredient of both the PBM and our model. As such, it is most likely not fully captured by a simple analytical formula. To summarise, our time delay prescription $\tau_{\rm d}$ depends linearly on $f_{\rm{err}}$, in addition to scaling as power laws with the general disc parameters $M_{\rm d}$, $j_{\rm{eff}}$ and $J_2$. Contrast this with the PBM, where the delays are effectively a function of the viscosity and accretion rate parameters that determine a solution of a specific accretion disc model.

An additional difference between our implementation and late iterations of the PBM are the latter's use of "time advances", which model the local warping of the disc towards the approaching secondary. In e.g. \citet{2018Dey}, these time advances are pre-computed by means of numerical simulations and are reported to amount to weeks and even months. The secondary black hole can have a strong local influence on the trajectory of a fluid element only when its gravitational pull is comparable to the primary's. For an impact at apoapsis the gravitational interaction time between a fluid element and the $\sim 100$ times lighter secondary will therefore be of the order of $\sim (1/2) (50\, r_{\rm{S}}/\sqrt{100})/(c/\sqrt{50}) \sim  20$ days, found by equating gravitational forces and dividing by typical orbital speeds at $\sim 50 \, r_{\rm{S}}$. By integrating the secondary's gravitational acceleration for an interaction time, the total out-of-plane displacement of the fluid element would amount to only a few AU, changing the impact timing by less than a day. {Large time advances must therefore be caused by more complex tidal interactions between the binary and the disc, which have been typically modelled via hydrodynamical simulations in the PBM literature. In the latter framework, they are considered an essential part of a precise timing model.} {In this work, we have found that the timing of optical flares can be fit with good precision without requiring any modelling of disc warping or time advances. It is plausible to attribute this major difference to our implementation of the disc's potential, which had been missing from previous iterations of the PBM. We do note that dynamical variations in the disc's scale height caused by tidal interactions are certainly physical and may be an improvement to our model's accuracy.} However, we do not wish to rely on numerical simulations of specific accretion discs, nor do our estimates suggest that they are an important contribution to the timing of flares at the target precision of $\sim 0.01$ yr. Most importantly, our analysis will also show that such advances are not required for our timing model to be predictive (see Section \ref{sec:sub:mock}).

\subsubsection{Relativistic Delays}
Several relativistic effects will influence the arrival times of radiation flares from massive systems at large distances \citep{1994karas,2010Dai}. The first and most obvious is cosmological red-shift, which simply stretches the total period by a factor $1 + z = 1.306$ \citep{1931lemaitre}.
In this work, we also implement the effect of Shapiro delays, $\tau_{\rm S}$, which produce impact radius dependent shifts of order weeks for our baseline parameters \citep{1964shapiro} and have also been neglected in the PBM:
\begin{align}
    \tau_{\rm S} \sim \frac{G M}{c^3}\log \left(\frac{D(z)}{r}\right) \sim 10 \text{ to }15 \text{ days,}
\end{align}
where $D$ is the distance corresponding to OJ-287's red-shift.
Römer time delays caused by the orientation of the disc could in principle contribute up to $\sim 50$ days if the disc were edge-on, the impact occurred at apo-apsis and the orbit's major axis were also aligned with the line of sight. Considering however that the disc is expected to be only $\sim 4^{\circ}$ off from a face-on configuration \citep{LV96}, the resulting shifts are suppressed by a small geometric factor, at most $\sim \sin(4^{\circ}) \sim 0.07$. Thus, they are typically close to or smaller than our target accuracy of $\sim 0.01$ yr. For the purposes of this work, we decide to neglect Römer delays in favor of having one less free parameter, which will allow us to fit our model to a subset of historical flare timings and "re-predict" the July 2019 flare (see Section \ref{sec:sub:mock}) without risking to over-fit the data.\newline \newline

\subsection{Fitting to the observed flare timings}
\label{sec:sub:fit}
\subsubsection{Parameter space and initial conditions}
To summarise, our updated flare timing model is determined by 7 parameters, three of which characterise the binary and four of which characterise accretion disc and its response to impacts (see Table \ref{tab:parameters}). Combined, these parameters represent a necessary compromise between the complexity required to properly capture the physics of OJ-287's system and the requirement to reduce the total number of degrees of freedom. In addition, we must specify the secondary's trajectory via several initial conditions. Since the timing of impacts is not affected by rotating the system along its symmetry axis, we can place the secondary in the mid-plane of the disc at the position $(-x_0,0,0)$ at an arbitrary initial time $t_0$. We choose to define the secondary's instantaneous velocity vector via a combination of its magnitude $v_0$, inclination $\Omega_0$ and tilt $\iota_0$ with respect to the z-axis (see Figure \ref{fig:sketch} and Table \ref{tab:parameters}). In total, a set of candidate flare timings thus depend on 7 system parameters and 4 initial conditions. The historical data provides us with 11 measurements (see Table \ref{tab:flares}). Additionally, we assume that there were indeed 19 flares between 1912 and 2019, some of which went undetected. This consideration adds an additional constraint that essentially fixes the orbital period, excluding some extreme orbital solutions that would account for large gaps between flares. Our model is therefore overdetermined, and we have the additional opportunity to only use a subset of the data, containing e.g. 10 flares (see Section \ref{sec:sub:mock}).

We must now devise a strategy to sample the 11-dimensional parameter space of our model and minimise the residuals between the candidate and the observed flare timings. In the PBM, the authors adopt what is essentially a manual minimisation routine, in which individual model parameters are selected a priori to roughly reproduce the data and satisfy basic astrophysical expectations. Further parameters are then adjusted one by one iteratively until every single flare occurs within its observed timing window \citep{LV96,2006valtonen,2018Dey,2022valtonen}. While this trial method can certainly produce valid solutions, it does not constitute a systematic search and gives no guarantee that the recovered parameters truly represent a global minimum. Most importantly, the resulting parameter uncertainties do not account for cross-correlations and degeneracies. They are therefore most likely vastly under-reported, even providing the validity of the underlying model. 

\begin{table}
    \centering
    \begin{tabular}{c|c|c}
      Parameters & Meaning & Initial prior    \\
      \hline
       $M$ & Primary mass & $\left[0.2 , 20\right]^*\times 10^{10}$ M$_{\odot}$  \\ 
       $ m $ & Secondary mass & $\left[0.1 , 10\right]^*\times10^{8}$ M$_{\odot}$ \\
       $\xi$ & Primary spin & $\left[0 , 1\right]$ \\
       $M_{\rm{d}}(100 r_{\rm{S}})$ & Disc scale mass & $\left[0.001 , 10\right]^*\times 10^{8}$ M$_{\odot}$  \\
       $j_{\rm{err}}$ & Disc profile & $\left[0,2\right]$ \\
       $-J_2$ & Disc quadrupole & $\left[0,\frac{1}{2}\right]$ \\
       $f_{\rm{err}}$ & Disc delay & $\left[0.001,1\right]^*$ \\
        & & \\
       Init. Condition & Meaning & Initial prior\\
       \hline
       $x_0$ & Position at $t_0$ & $\left[3 , 100\right]^*\times $ $r_{\rm S}$ \\
       $v_0$ & Speed at $t_0$ & $\left[0.01 , 1\right]^*\times$ $c$\\
       $\Omega_0$ & Inclination at $t_0$ & $\left[0 , 2 \pi\right]$  \\
       $\iota_0$ & Tilt at $t_0$ & $\left[-\frac{\pi}{2}, \frac{\pi}{2}\right]$ 
    \end{tabular}
    
    \caption{Meaning and initial priors of the 7 parameters and 4 initial conditions required to produce a set of trial flare timings. The priors are taken to be uniform in the reported range, or log-uniform if denoted by an asterisk. Note than the large majority of parameters drawn from these wide priors lead to wildly incorrect total number of flares within the observation time.}
  
    \label{tab:parameters}
\end{table}

\subsubsection{MCMC sampling and numerical pipeline}
MCMC sampling provides a framework in which both of the issues mentioned above are addressed naturally. For the purposes of this work, we use the MCMC sowtware \texttt{emcee} \citep{2013emcee} which supports efficient parallelisation via the \texttt{multiprocess} package \citep{2012multi}. We adopt a standard likelihood function $\mathcal{L}$ of the form:
\begin{align}
   \log \left( \mathcal{L}\right) = - \frac{1}{2} \sum \frac{\left(\text{trial} - \text{data} \right)^2 }{\text{uncertainty}^2}
\end{align}
where we attempt to minimize the least-squares difference between some trial flare timings and the historical data, while consistently accounting for the various data gaps. Additionally, we enforce the correct total number of flares by returning a vanishing likelihood when the appropriate condition is not met. To recover the parameter posteriors, we initially run  64 MCMC walkers on wide uninformed priors reported in Table \ref{tab:parameters}. After approximately $\sim 15'000$ iterations, the walkers have thoroughly scouted parameter space and identified all likelihood maxima. In order to re-sample the most interesting part of the posterior, we then re-initialise 32 walkers in a small neighbourhood of the global likelihood maximum, running further MCMC chains until the Gelman-Rubin criterion is met \citep{1992gelman,2018vats} and the best fit values are converged within their standard deviations. Convergence typically requires approximately $\sim 30'000$ iterations.\newline \newline
For the purposes of this work, we ran the numerical pipeline with a data-set containing all 11 historical flare timings, forming the basis of our results in Sections \ref{sec:sub:posteriors} and \ref{sec:sub:predictions}. Additionally, we re-ran it on a data set which excludes the July 2019 flare, in order to confirm the predictive power of our model (see Section \ref{sec:sub:mock}). In total, we evaluated just under 4'000'000 trial orbits\footnote{Compare this with the order thousand evaluated trial orbits as reported in e.g. \citet{2018Dey}}, for a total of $\sim 850$ cpu hours. Thanks to the magic of parallelisation, the whole computation was performed on a 8-core Lenovo laptop over the course of several weeks.
\begin{figure*}
    \centering
    \includegraphics[scale=0.3]{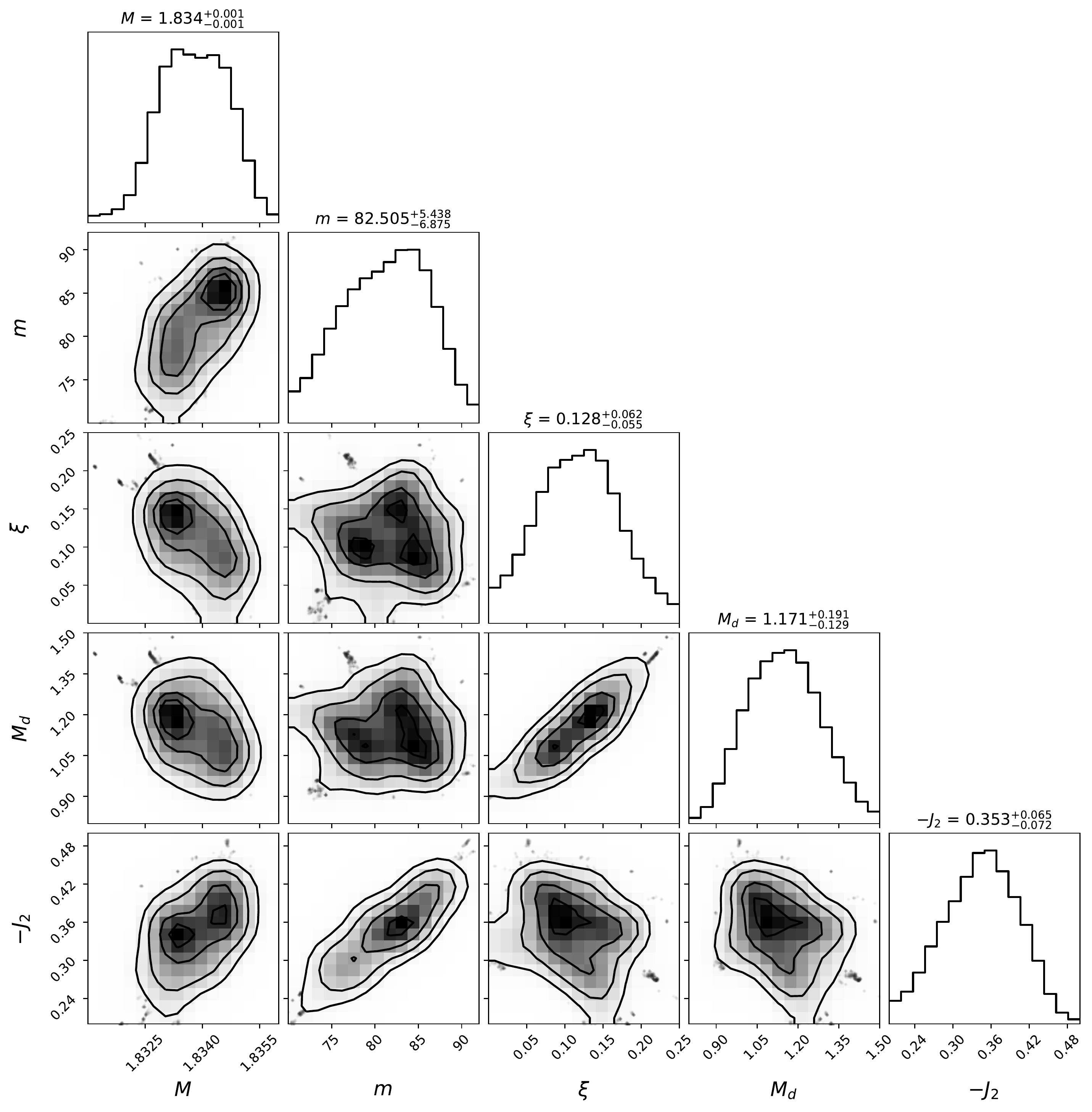}
    \includegraphics[scale=0.3]{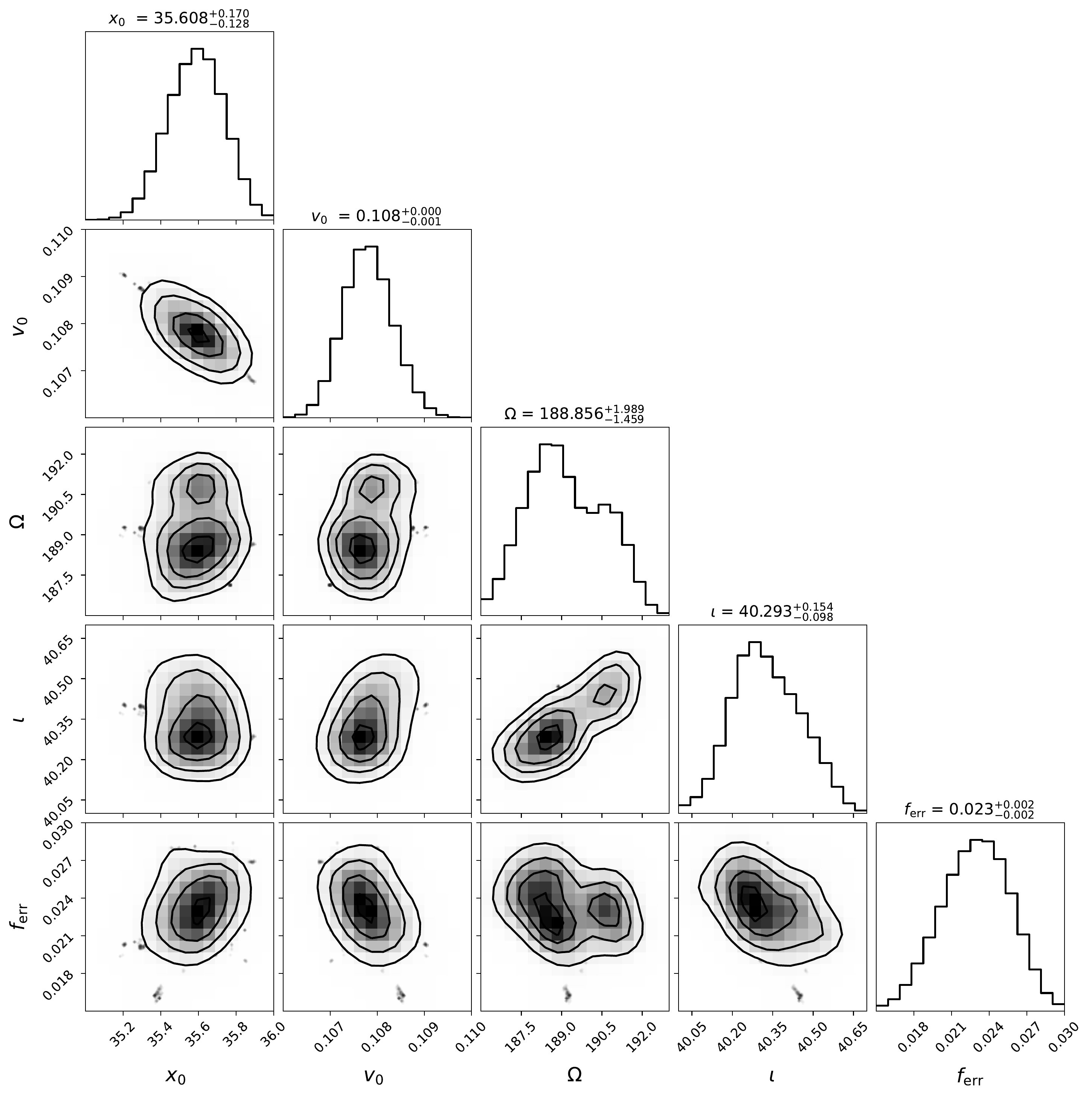}
    
    \includegraphics[scale=0.54]{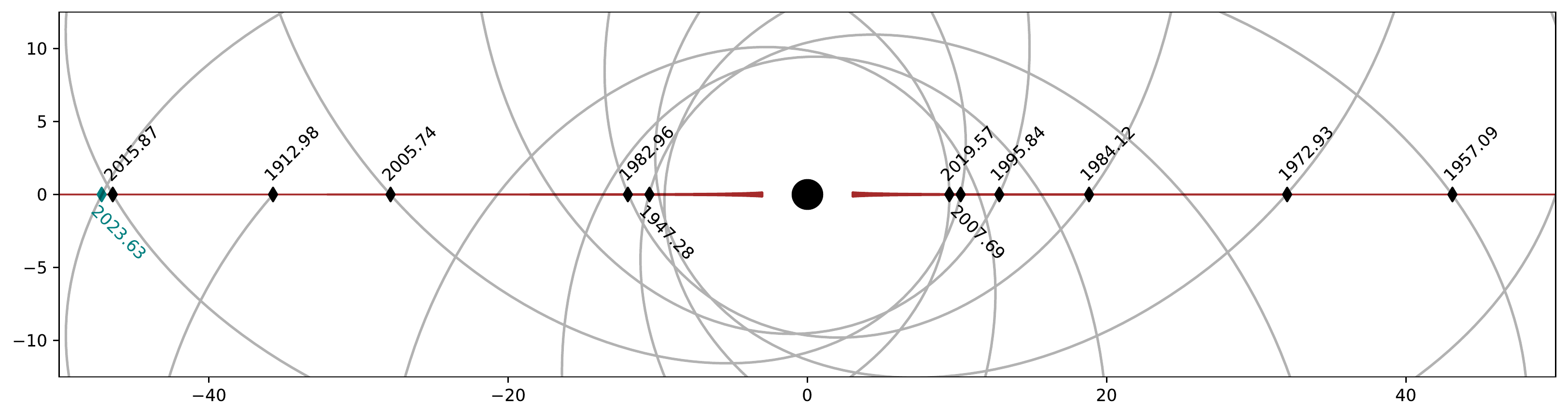}
    \caption{The two top panels show the posteriors of the binary and disc parameters, as well as the initial conditions recovered after approximately 30'000 iterations of an MCMC run with 32 walkers, initialised in the vicinity of the best-fit value.  Note the bi-modality in the initial inclination $\Omega_0$, and its effect on the marginalised posterior of e.g. the primary mass $M$. Note also the correlations in the disc's mass scale and primary spin, as well as the disc's quadrupole moment $J_2$ and the secondary's mass $m$. The lower panel shows a to-scale visualisation of the best-fit disc and orbit solution for OJ-287's system. The diamond markers denote the position of past impacts (black) that have led to a detected optical flare. The teal marker denotes the prospective 26th flare on the 21st of August 2023 $\pm$ 32 days.}
    \label{fig:posteriors}
\end{figure*}

\begin{table}
    \centering
    \begin{tabular}{c|c|c}
   Model components  & This work & PBM    \\
      \hline
       Binary EoM & 3 PN & 3.5 PN + PN tail\\ 
       Disc EoM & Yes & \textit{No!} \\
       Disc structure & Power law & $\alpha$-disc \\
       Time delays & As in PBM & As in PBM \\
       Shapiro delays & Yes & No \\
       Römer delay & No & Yes \\
       Time advances & No & Yes \\

       Fitting method & MCMC & trial method \\
       Nr. trial orbits & $\sim 10^6$ & $\sim 10^3$ \\
    \end{tabular}
    
    \caption{Differences in dynamical ingredients and methodology between our model and the PBM. Highlighted is the inclusion of the gravitational influence of the accretion disc, which is ultimately responsible for the large timing shifts discussed in section \ref{sec:sub:predictions}. An important excercise, planned for future work, is to precisely quantify the effect of every individual modelling choice on the timing, beyond the qualitative arguments presented throughout Section \ref{sec:methods} \citep[in the vein of e.g.][]{1998Pietila}.}
  
    \label{tab:model}
\end{table}

\section{Results}
\label{sec:results}
\subsection{Binary and disc parameter posteriors}
\label{sec:sub:posteriors}
\subsubsection{General features and correlations}
The results of our MCMC run including all 11 historical flare timings are a set of posterior distributions for the 7 parameters and 4 initial conditions of our model. They are visualised in Figure \ref{fig:posteriors} as two corner plots \citep{corner}. The best fit values and the uncertainties resulting from the marginalised posteriors are listed in Table \ref{tab:posteriors} and compared to the most up-to-date results of the PBM. In general, the recovered binary and disc parameters qualitatively reproduce the broad expectations set by the framework of the PBM. However, they do differ in several important details. Here we comment on the ones that shed most light on the physics of the system.

Firstly, there is a clear bi-modality in the posterior for the initial inclination $\Omega_0$. The two likelihood peaks are separated by around $3^{\circ}$ around a mean value of $\sim 189.5^{\circ}$. The common PBM assumption that the orbit be perfectly perpendicular to the disc plane is excluded, a constraint that arises from properly modelling the effects of a preferential symmetry plane, i.e. the disc's gravitational quadrupole, on the flare timing. The bi-modality in $\Omega_0$ is also responsible for widening the posteriors of many other parameters, including the primary mass $M$. This leads to factor 10 larger uncertainty with respect to the PBM result, despite the almost identical recovered best-fit value of $1.834\times 10^{10}$ M$_{\odot}$. Additionally, we recover a significantly lower value for the primary's spin with respect to the PBM. Its posterior distribution shows a clear correlation with the disc's scale mass, which arises due to an interesting interaction with the disc's monopole graviational moment. The latter forces a precession-like effect at apoapsis, where the enclosed mass is large, rather than at periapsis where relativistic effects (such as spin induced frame dragging) are large. Furthermore, the correlation significantly broadens the range of likely spin values, leading to an uncertainty of $\sim 50\%$ of the best-fit value rather than the remarkable (and perhaps over-optimistic) $\sim 1\%$ reported in the PBM. The best fit value for the secondary mass $m$ is $0.82 \times 10^8 $ M$_{\odot}$ approximately half of the reported result in the PBM. This large discrepancy is related to the introduction of the disc's quadrupole moment and there are hints of a correlation between the parameters $m$ and $J_2$. Physically, this behaviour arises from small offsets in the system's centre of mass being degenerate with spurious quadrupolar contributions to the potential.

\subsubsection{Dynamical constraints on OJ-287's accretion disc}
Provided that our EoM are correct and that our delay prescriptions are accurate, our work provides a novel measurement of a quasar accretion disc's mass profile and quadrupole moment that is not based on specific disc models, luminosity scaling relations or spectral data. We recover a disc mass scale $M_{\rm d}$ at $100 \,r_{\rm S}$ of $1.2\times 10^8$ M$_{\rm{\odot}}$, approximately 5 times lower than the best fit values in the PBM \citep[which we extrapolate from][]{2019valtonendisc}. Within the sampled radii of $\sim 10$ to $\sim 50 $ $r_{\rm{S}}$, the disc's enclosed mass grows approximately linearly, scaling as $r^{2 - j_{\rm{eff}}}$, where the best fit value is given by $j_{\rm{eff}} = 0.92 \pm 0.12$. As mentioned previously, in the limit of thin discs the quadrupole moment $J_2$ and $j_{\rm{eff}}$ are related by a simple formula, given in Eq. \ref{eq:jeff}. In our analysis, this turns out (a posteriori) to be an extremely good approximation, confirming that OJ-287's accretion disc can be indeed modelled as thin, at least gravitationally. The individual parameters $j_1$ and $j_2$ describing the density and scale height profiles are therefore only very poorly constrained. This suggests that in effect our disc model only truly depends on two independent free parameters rather than three.

Interestingly, the flare timing data selects a disc with profiles roughly following the expectations of the PBM $\alpha$-disc, despite our model exclusively relying on dynamical information. {This can bee seen as a strong validation of using such a disc in the first place, provided that the underlying model is correct}. Indeed, the recovered disc in \citet{2019valtonendisc} is shown to have an approximately constant scale height and a decreasing density, from which we can estimate values of $j_{\rm eff} \sim 0.6$ and $J_2\sim 0.4$. Given our setup, we cannot directly estimate an accretion rate nor a viscosity from our recovered parameters. However, assuming for simplicity the same viscous timescale as in the PBM, the factor 5 reduction of the total mass budget more easily aligns with the luminosity and spectral constraints highlighted recently in \citet{2023komossa} and \citet{2023komossa2}. This additionally showcases the advantage of not forcing the model to fit an $\alpha$-disc, for which the aforementioned observational constraints are inconsistent with such a large primary mass. Finally, the disc delay coefficient takes a best fit value of $\sim 0.02$, leading to delays of order days to months, also compatible with previous expectations from the PBM. {Despite the many similarities, it is important to stress that our disc model is fundamentally different than the one used in the PBM, as it does not allow for warps and local scale height variations that cause timing advances. The fact that, with our model, the flare timings may be fitted without requiring such elements is a further indication of the importance of modelling the accretion disc's gravitational potential. However, the assumption of a rigid disc is certainly a simplification, and is likely reflected in the best fit posterior parameters. Our disc is best understood as being an effective model, which suffices for the purposes of timing the optical flares. Determining whether this disc model can explain all other electromagnetic signatures of OJ-287 is significantly beyond the scope of this work, but an interesting avenue for further investigation.}

For visualisation purposes, a to-scale rendition of the best-fit binary and disc components comprising of OJ-287's system is shown in Figure \ref{fig:posteriors}, along with the trajectory of the secondary over the last century.

\begin{table}
    \centering
    \begin{tabular}{c|c|c|c}
      Parameter & Unit & This work  & PBM  \\
      \hline
       $M$ & $10^{10}$ M$_{\odot}$ &$1.834 \pm 0.001 $ & $1.8348 \pm 0.0008 $      \\
       $m$ & $10^{8}$ M$_{\odot}$  & $0.82 \pm 0.06 $ & $1.50 \pm 0.03$     \\
       $\xi$ & - & $0.13 \pm 0.06$ &   $0.381 \pm 0.004$    \\
       $M_{\rm d}(100\, r_{\rm S})$ & $10^8$ M$_{\odot}$ & $1.2 \pm 0.2$  & est. $\sim 5.1  $   \\
       $j_{\rm {eff}}$ & - & $0.92\pm 0.12$  & est. $\sim 0.6  $\\
       $-J_2$ & -  &$0.35 \pm 0.07$ &  est. $\sim 0.4  $ \\    
    \end{tabular}
    
    \caption{The best fit parameters and their uncertainties in our model versus the reported ones in the PBM. For the sake of comparison, we estimate the PBM values for the disc parameters $M_{\rm d}$, $j_{\rm{eff}}$ and $J_2$ from the results reported in \citet{2019valtonendisc}, since they are not fitted directly. Beyond the primary mass, none of the recovered values overlap, re-affirming how one must always take such constraints with caution, as they rely on the assumption that the underlying model is indeed the correct one.}
    
    \label{tab:posteriors}
\end{table}

\subsection{Predictions and uncertainties for future flares}
\label{sec:sub:predictions}
\begin{figure}
    \centering
    \includegraphics[scale=0.69]{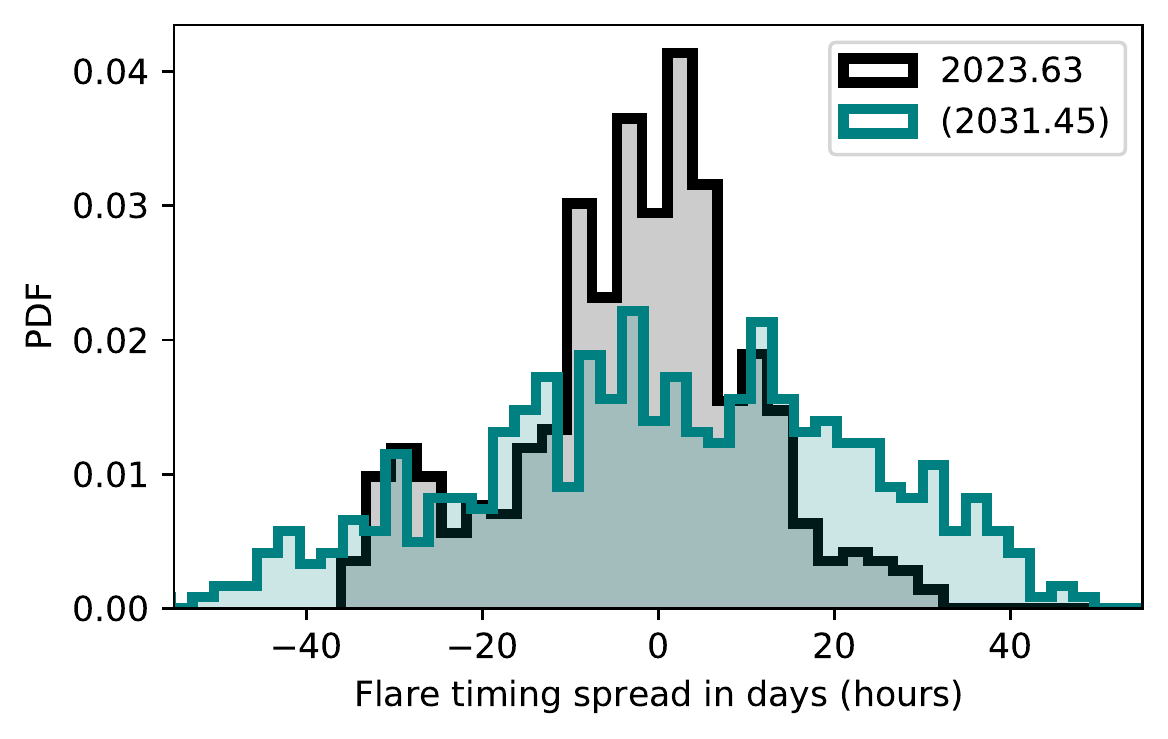}
    \includegraphics[scale=0.7]{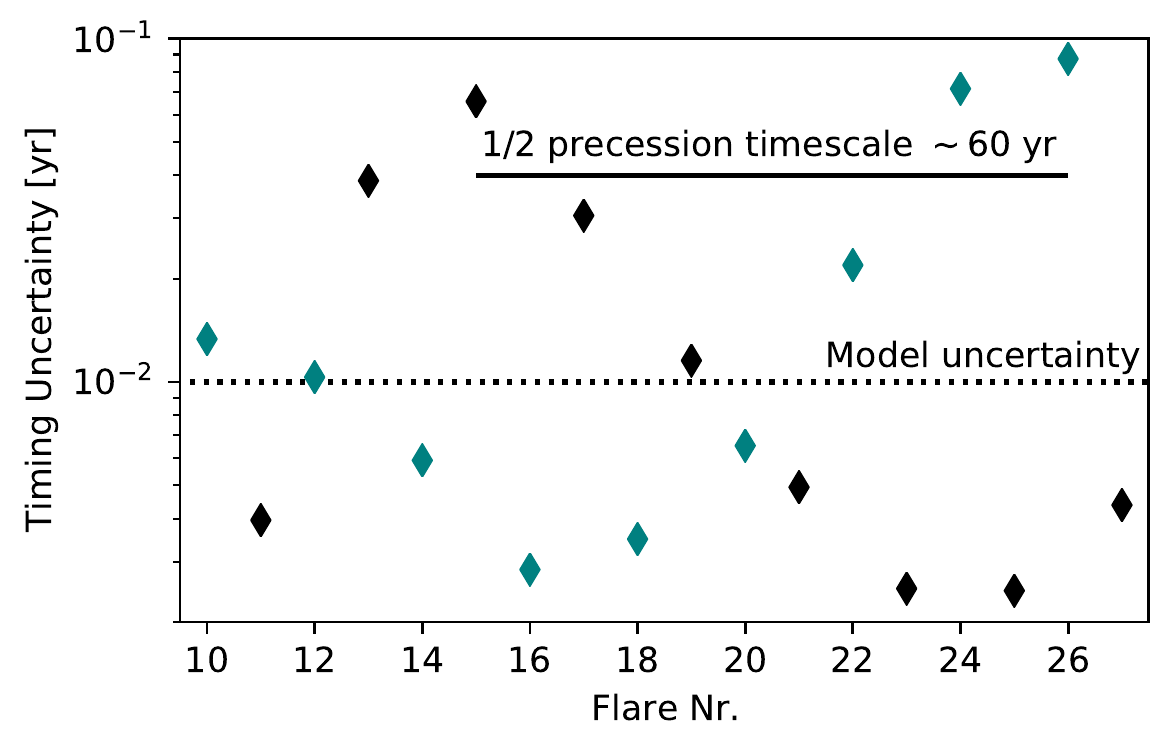}
    
    \caption{Top panel: we compare the PDF for the 26th and 27th impact flares according to our model, centered on their expected dates of the 21st August 2023 and the 16th of July 2031 (plotted with different scales). Note how the multi-modality present in the parameter posteriors is reflected in the expected timing. Bottom panel: we show the uncertainties of several past and future flares, uncovering a very clear alternating modulated pattern that can be associated to the relativistic precession timescale of 120 yr. We also highlight the uncertainty arising from our model's limitations.}
    \label{fig:spread}
\end{figure}
In stark contrast to the most recent PBM claims, our model predicts that the next flare, i.e. the 26th in the customary numbering system, should occur on the \textbf{21st of August 2023} $\pm$ \textbf{32 days}. The expected date is thus shifted by by $\sim$13 months with respect to \citep{2023valtonen}. This large discrepancy is entirely caused by the addition of the gravitational influence of the disc in the EoM and the resulting differences in the recovered parameter posteriors. Indeed, comparing the best fit trajectory in Figure \ref{fig:posteriors} with e.g. Figure 3 in \citet{2018Dey}, our model forecasts that the location of the 26th impact will also differ drastically from previous expectations, being much closer to the site of the 2015 impact. If detected, the upcoming flare is therefore likely to have similar characteristics to the well studied 2015 Centenary flare when it comes to its luminosity, duration and spectrum \citep{2015komossa,2015ciprini,2015shappe,2016valtonen,2019valtonendisc}. Beyond the epoch itself, the uncertainty in our prediction is much larger than the typical timing uncertainties that have been reported in the PBM. While this is partially explained by the wider posteriors in our recovered parameters (see Table \ref{tab:posteriors}), further investigation reveals some interesting details.

In the top panel of Figure \ref{fig:spread}, we construct the probability distribution function (PDF) of the 26th flare by drawing 500 posterior samples and collecting the timing results in a histogram. There is clear evidence that the multi-modality present in the parameter posteriors (specifically the inclination $\Omega_0$) is carried over in the timing PDF, which shows a primary peak accompanied by at least an additional local maximum shifted by approximately -30 days. In similar fashion, we compute the standard deviation in the timing of several past and future flares, and plot the results in the bottom panel of Figure \ref{fig:spread}. We uncover a very clear alternating pattern, in which each consecutive flare is characterised by an unusually large or small uncertainty. Indeed, the uncertainty for the 27th flare on the 16th of July 2031 is only a few days, as is also shown in the top panel of Figure \ref{fig:spread}. Furthermore, the pattern presents a modulation on a timescale of roughly $60$ years, clearly associated to the system's relativistic perihelion precession. The latter causes the system's orientation to rotate fully in approximately 120 yr, forcing the trajectory of the secondary to extrude primarily above or below the accretion disc's plane for roughly half as much time. The alternating pattern in the timing uncertainty is also associated to the obit's orientation with respect to the disc. By inspecting the secondary's trajectory, we observe that impacts for which the orbital nodes are aligned with the minor axis are characterised by either large (months) or small uncertainties (days). Conversely, impacts for which the nodes are aligned with the major axis typically present average uncertainties (weeks). Thus, the exact timing of the upcoming flare unfortunately happens to be the most uncertain of the last century, while the prospective 16th of July 2031 flare is much more easily constrained, having an uncertainty of only $\sim 40$ hr. Note that both the timing PDF multi-modality and the alternating pattern in the uncertainties are features that are cleary associated with the real physics of the system. They could only be uncovered with a sophisticated sampling of the posterior distributions, which in this work has been accomplished via MCMC methods.

{We note that recent investigations of OJ-287's light curve have revealed evidence for the secondary's impact with the disc in early 2022, in the form of a blue flare \citep{2023valtonendiss}. If it is indeed produced by an impact, the observation of such a flare clearly invalidates both the results of our parameter recovery and our predictions for future flares. It does not however invalidate the requirement of any precessing binary model to properly include the gravitational potential of the accretion disc, which can produce timing shifts of order 1 yr. Furthermore, we have shown how a sophisticated sampling technique, e.g. MCMC sampling, is required to truly understand the intricacies of the timings. We look forward to incorporating additional information, e.g. from additional flares, and revisiting elements of our models (listed in Table~\ref{tab:model}) in future iterations of this work.}

{It is also the case that, at the time of revising this manuscript (22/09/2023), Earth based tracking of OJ-287's light curve has recommenced after the yearly period (early July to early September) in which the Sun's position affects observations. An up-to date visualisation of the system's light curve is curated online \footnote{https://www.as.up.krakow.pl/sz/oj287.html}. It can be seen that observations via the KRK instrument of the Jagellonian University recommence in the early days of September. They show hints of a preceding luminosity maxima, somewhere around mid August, that may have occurred during the observational gap. However, with the given data it is clearly impossible to determine whether an optical flare may or may not have actually taken place in the period predicted by our work, at least by naive analyisis. This is unfortunate, since the timing of the 26th flare sets a clear distinction between our work and the predictions of the PBM. Nevertheless, it is interesting in its own right that both models instead give very similar results for the previous optical flare, despite the very different dynamical and methodological ingredients.}

\subsection{Mock prediction of the 2019 Eddington flare}
\label{sec:sub:mock}
\begin{figure}
    \centering
    \includegraphics[scale=0.69]{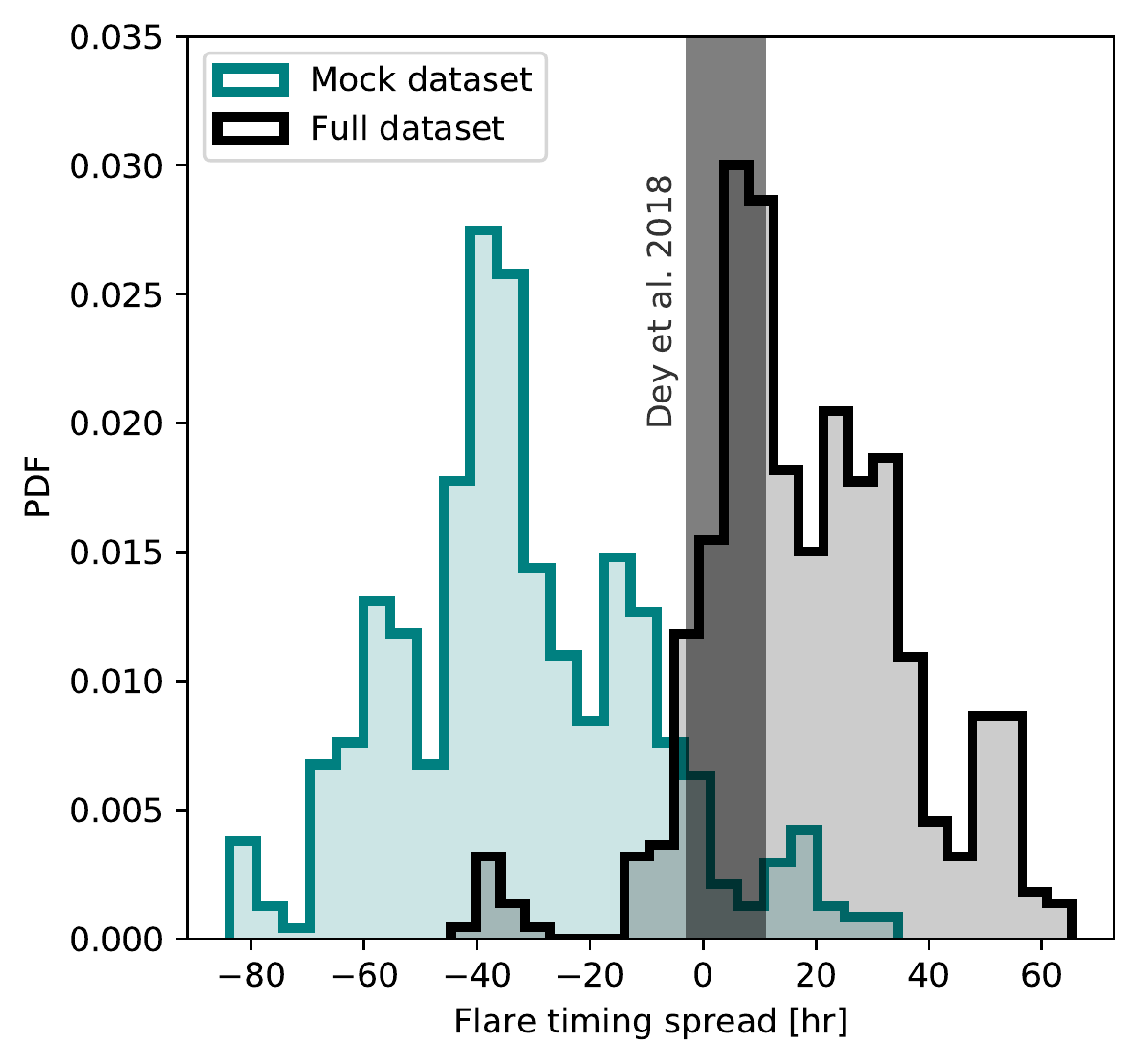}
    \caption{We compare the PDFs for the 25th impact flare epoch with the observed timing, normalised to the value 0 \citep{2020Laine}. The teal coloured distribution is conditioned on a mock data-set including only data available prior to 2019, showcasing the predictive power of our model. For comparison, we also show the timing distribution resulting from the full data-set (grey), which by construction peaked around the observed timing. The remarkable prediction by \citet{2018Dey} is denoted by the grey bar, {where the timing has been extracted from \citet{2020Laine}.}}
    \label{fig:2019}
\end{figure}
As discussed thoroughly in Section \ref{sec:sub:fit}, our model only requires 10 historical flare timings to be completely determined. Therefore, we are able run our numerical pipeline on a subset of the historical timing data that does not contain the July 2019 flare, in order to test whether our model would have been able to predict the latter with comparable accuracy to \citet{2018Dey} and \citet{2020Laine}.
Once again, the result of our numerical runs are a set of parameter posteriors, differing only slightly from what is shown in Figure \ref{fig:posteriors}. While uncertainties are generally increased due to more pronounced correlations, the new posteriors still preserve the patterns in the timing uncertainties discussed in Section \ref{sec:sub:predictions}. Thus, conditioning our model to this mock dataset leads to a relatively well constrained "mock prediction" for the July 2019 flare.
Figure \ref{fig:2019} shows the PDF of the 25th impact flare, computed by collecting the timing results from 500 draws of the new posterior distributions. According to the PDF, our model would have been able to predict the Eddington flare within 40 hr of the actual confirmed detection \citep{2020Laine}. The timing PDF is peaked around the 29th of July $\pm 33$ hr and overlaps with the actual detection epoch.  

\section{Conclusion}
\label{sec:conclusion}
In this paper, we have constructed an alternative precessing binary model to explain the quasi-periodic optical flares in OJ-287's light curve. In contrast to the model first proposed by \citet{LV96} (and then iterated upon over many years), our work consistently accounts for the gravitational influence of the accretion disc by decomposing it into a monopole and a quadrupole contribution. Furthermore, we have shown how sophisticated sampling techniques are required to uncover realistic correlations and uncertainties in both parameter posteriors and flare timings. The results of our work are discussed throughout Section \ref{sec:results}, and compared with the established literature. However, it is important to highlight the predictions of our model which { may differ from previous expectations}:
\begin{itemize}
    \item Our model can reproduce the timing of the July 2019 flare within 40 hr of it's detection, when conditioned with timing data only available prior to it.

    \item In stark contrast with established literature, our model predicts that the 26th optical flare should have occured on the 21st of August 2023 $\pm$ 32 days. {Unfortunately, the majority of this range lies in the period when OJ-287 is unobservable to Earth based telescopes.}

    \item The location of the 26th impact is extremely close to the previous impact responsible for the 2015 Centenary flare. The flare is therefore likely to have similar luminosity, duration and spectral properties. 
\end{itemize}
The crucial insight however, is that despite the very different ingredients, both the PBM and our model have (or would have) been able to successfully predict the timing of the great 2019 Eddington flare. {The inclusion of the accretion disc's potential makes the binary model predictive without requiring the complex numerical modelling of disc and its response to the perturber as required in the PBM}. However, from the pattern revealed in the timing uncertainties (see Section \ref{sec:sub:predictions} and Figure \ref{fig:spread}), we have shown that the epoch of the 25th flare is incredibly robust to changes in the system's parameters. By analogy, it must also be robust to changes in the chosen underlying model. {This clearly indicates how the simpler} framework for a precessing black hole binary {adopted in this work is able to explain and re-predict past flare timings} in OJ-287's light curve. Additionally, it validates the prospect of using flare timing data to directly measure the gravitational potential of quasar accretion discs (see Section \ref{sec:sub:posteriors}), revealing their structure and composition in an unprecedented way. What is perhaps more difficult to interpret is the difference in the predicted epoch for the 26th flare. {Within the assumptions of our model,} including the dynamical influence of the accretion disc causes a timing shift of almost a full year with respect to previous expectation. Interestingly, the shift occurs despite an almost identical recovered primary black hole mass. {Clearly, only a thorough investigation of the differences between the PBM and our own model can help shed light on the truly essential elements of the precessing binary framework}. Another possibility is that the characteristics of OJ-287's light curve are conclusively shown not to be associated with a binary system. Then it would be time to revise and refine alternative frameworks, such that they may become predictive rather than only descriptive.  To be able to conclusively answer these questions, we would like to strongly encourage the observational community to keep tracking this fascinating system over the months of August and September and the upcoming years.

\section*{Acknowledgements}
The authors acknowledge support from the Swiss National Science Foundation under the Grant 200020\_192092. LZ acknowledges Pedro R. Capelo, Mudit Garg and Andrea Derdzinski for several helpful discussions. LZ acknowledges the institution and participants of GALSTAR UZH.

\section*{Data availability}
The data underlying this article will be shared on reasonable request to the authors.

\scalefont{0.94}
\setlength{\bibhang}{1.6em}
\setlength\labelwidth{0.0em}
\bibliographystyle{mnras}
\bibliography{main}

\begin{thebibliography}{}
\makeatletter
\relax
\def\mn@urlcharsother{\let\do\@makeother \do\$\do\&\do\#\do\^\do\_\do\%\do\~}
\def\mn@doi{\begingroup\mn@urlcharsother \@ifnextchar [ {\mn@doi@}
  {\mn@doi@[]}}
\def\mn@doi@[#1]#2{\def\@tempa{#1}\ifx\@tempa\@empty \href
  {http://dx.doi.org/#2} {doi:#2}\else \href {http://dx.doi.org/#2} {#1}\fi
  \endgroup}
\def\mn@eprint#1#2{\mn@eprint@#1:#2::\@nil}
\def\mn@eprint@arXiv#1{\href {http://arxiv.org/abs/#1} {{\tt arXiv:#1}}}
\def\mn@eprint@dblp#1{\href {http://dblp.uni-trier.de/rec/bibtex/#1.xml}
  {dblp:#1}}
\def\mn@eprint@#1:#2:#3:#4\@nil{\def\@tempa {#1}\def\@tempb {#2}\def\@tempc
  {#3}\ifx \@tempc \@empty \let \@tempc \@tempb \let \@tempb \@tempa \fi \ifx
  \@tempb \@empty \def\@tempb {arXiv}\fi \@ifundefined
  {mn@eprint@\@tempb}{\@tempb:\@tempc}{\expandafter \expandafter \csname
  mn@eprint@\@tempb\endcsname \expandafter{\@tempc}}}

\bibitem[\protect\citeauthoryear{{Abramowicz} \& {Fragile}}{{Abramowicz} \&
  {Fragile}}{2013}]{2013abramowicz}
{Abramowicz} M.~A.,  {Fragile} P.~C.,  2013, \mn@doi [Living Reviews in
  Relativity] {10.12942/lrr-2013-1}, \href
  {https://ui.adsabs.harvard.edu/abs/2013LRR....16....1A} {16, 1}

\bibitem[\protect\citeauthoryear{{Antonucci}}{{Antonucci}}{1993}]{1993antonucci}
{Antonucci} R.,  1993, \mn@doi [\araa] {10.1146/annurev.aa.31.090193.002353},
  \href {https://ui.adsabs.harvard.edu/abs/1993ARA&A..31..473A} {31, 473}

\bibitem[\protect\citeauthoryear{{Barausse}}{{Barausse}}{2012}]{2012barausse}
{Barausse} E.,  2012, \mn@doi [\mnras] {10.1111/j.1365-2966.2012.21057.x},
  \href {https://ui.adsabs.harvard.edu/abs/2012MNRAS.423.2533B} {423, 2533}

\bibitem[\protect\citeauthoryear{{Barker} \& {O'Connell}}{{Barker} \&
  {O'Connell}}{1975}]{1975barker}
{Barker} B.~M.,  {O'Connell} R.~F.,  1975, \mn@doi [\prd]
  {10.1103/PhysRevD.12.329}, \href
  {https://ui.adsabs.harvard.edu/abs/1975PhRvD..12..329B} {12, 329}

\bibitem[\protect\citeauthoryear{{Barker}, {O'Brien}  \& {O'Connell}}{{Barker}
  et~al.}{1981}]{1981barker}
{Barker} B.~M.,  {O'Brien} G.~M.,   {O'Connell} R.~F.,  1981, \mn@doi [\prd]
  {10.1103/PhysRevD.24.2332}, \href
  {https://ui.adsabs.harvard.edu/abs/1981PhRvD..24.2332B} {24, 2332}

\bibitem[\protect\citeauthoryear{{Bernard}, {Blanchet}, {Faye}  \&
  {Marchand}}{{Bernard} et~al.}{2018}]{2018bernard}
{Bernard} L.,  {Blanchet} L.,  {Faye} G.,   {Marchand} T.,  2018, \mn@doi
  [\prd] {10.1103/PhysRevD.97.044037}, \href
  {https://ui.adsabs.harvard.edu/abs/2018PhRvD..97d4037B} {97, 044037}

\bibitem[\protect\citeauthoryear{{Binney} \& {Tremaine}}{{Binney} \&
  {Tremaine}}{1987}]{1987binney}
{Binney} J.,  {Tremaine} S.,  1987, {Galactic dynamics}

\bibitem[\protect\citeauthoryear{{Blanchet}}{{Blanchet}}{2014}]{2014blanchet}
{Blanchet} L.,  2014, \mn@doi [Living Reviews in Relativity]
  {10.12942/lrr-2014-2}, \href
  {https://ui.adsabs.harvard.edu/abs/2014LRR....17....2B} {17, 2}

\bibitem[\protect\citeauthoryear{{Bondi}}{{Bondi}}{1952}]{1952Bondi}
{Bondi} H.,  1952, \mn@doi [\mnras] {10.1093/mnras/112.2.195}, \href
  {https://ui.adsabs.harvard.edu/abs/1952MNRAS.112..195B} {112, 195}

\bibitem[\protect\citeauthoryear{{Britzen} et~al.,}{{Britzen}
  et~al.}{2018}]{2018Britzen}
{Britzen} S.,  et~al., 2018, \mn@doi [\mnras] {10.1093/mnras/sty1026}, \href
  {https://ui.adsabs.harvard.edu/abs/2018MNRAS.478.3199B} {478, 3199}

\bibitem[\protect\citeauthoryear{{Browne}}{{Browne}}{1971}]{1971Browne}
{Browne} I.~W.~A.,  1971, \mn@doi [\nat] {10.1038/231515a0}, \href
  {https://ui.adsabs.harvard.edu/abs/1971Natur.231..515B} {231, 515}

\bibitem[\protect\citeauthoryear{{Butuzova} \& {Pushkarev}}{{Butuzova} \&
  {Pushkarev}}{2020}]{2020butozova}
{Butuzova} M.~S.,  {Pushkarev} A.~B.,  2020, \mn@doi [Universe]
  {10.3390/universe6110191}, \href
  {https://ui.adsabs.harvard.edu/abs/2020Univ....6..191B} {6, 191}

\bibitem[\protect\citeauthoryear{{Carangelo}, {Falomo}, {Kotilainen}, {Treves}
  \& {Ulrich}}{{Carangelo} et~al.}{2003}]{2003Carangelo}
{Carangelo} N.,  {Falomo} R.,  {Kotilainen} J.,  {Treves} A.,   {Ulrich} M.~H.,
   2003, \mn@doi [\aap] {10.1051/0004-6361:20031519}, \href
  {https://ui.adsabs.harvard.edu/abs/2003A&A...412..651C} {412, 651}

\bibitem[\protect\citeauthoryear{{Chen}, {Jiang}, {Goodman}  \&
  {Ostriker}}{{Chen} et~al.}{2023}]{2023chen}
{Chen} Y.-X.,  {Jiang} Y.-F.,  {Goodman} J.,   {Ostriker} E.~C.,  2023, \mn@doi
  [arXiv e-prints] {10.48550/arXiv.2302.10868}, \href
  {https://ui.adsabs.harvard.edu/abs/2023arXiv230210868C} {p. arXiv:2302.10868}

\bibitem[\protect\citeauthoryear{{Ciprini}, {Perri}, {Verrecchia}  \&
  {Valtonen}}{{Ciprini} et~al.}{2015}]{2015ciprini}
{Ciprini} S.,  {Perri} M.,  {Verrecchia} F.,   {Valtonen} M.,  2015, The
  Astronomer's Telegram, \href
  {https://ui.adsabs.harvard.edu/abs/2015ATel.8401....1C} {8401, 1}

\bibitem[\protect\citeauthoryear{{Corso}, {Purcell}, {Giroux}  \&
  {Schultz}}{{Corso} et~al.}{1984}]{1984Corso}
{Corso} G.~J.,  {Purcell} B.,  {Giroux} M.,   {Schultz} J.,  1984, \mn@doi
  [\pasp] {10.1086/131408}, \href
  {https://ui.adsabs.harvard.edu/abs/1984PASP...96..705C} {96, 705}

\bibitem[\protect\citeauthoryear{{Craine} \& {Warner}}{{Craine} \&
  {Warner}}{1973}]{1973Craine}
{Craine} E.~R.,  {Warner} J.~W.,  1973, \mn@doi [\apjl] {10.1086/181115}, \href
  {https://ui.adsabs.harvard.edu/abs/1973ApJ...179L..53C} {179, L53}

\bibitem[\protect\citeauthoryear{{Dai}, {Fuerst}  \& {Blandford}}{{Dai}
  et~al.}{2010}]{2010Dai}
{Dai} L.~J.,  {Fuerst} S.~V.,   {Blandford} R.,  2010, \mn@doi [\mnras]
  {10.1111/j.1365-2966.2009.16038.x}, \href
  {https://ui.adsabs.harvard.edu/abs/2010MNRAS.402.1614D} {402, 1614}

\bibitem[\protect\citeauthoryear{{Dey} et~al.,}{{Dey} et~al.}{2018}]{2018Dey}
{Dey} L.,  et~al., 2018, \mn@doi [\apj] {10.3847/1538-4357/aadd95}, \href
  {https://ui.adsabs.harvard.edu/abs/2018ApJ...866...11D} {866, 11}

\bibitem[\protect\citeauthoryear{{Dey}, {Valtonen}, {Gopakumar}, {Lico},
  {G{\'o}mez}, {Susobhanan}, {Komossa}  \& {Pihajoki}}{{Dey}
  et~al.}{2021}]{2021Dey}
{Dey} L.,  {Valtonen} M.~J.,  {Gopakumar} A.,  {Lico} R.,  {G{\'o}mez} J.~L.,
  {Susobhanan} A.,  {Komossa} S.,   {Pihajoki} P.,  2021, \mn@doi [\mnras]
  {10.1093/mnras/stab730}, \href
  {https://ui.adsabs.harvard.edu/abs/2021MNRAS.503.4400D} {503, 4400}

\bibitem[\protect\citeauthoryear{{Dunlop}, {McLure}, {Kukula}, {Baum}, {O'Dea}
  \& {Hughes}}{{Dunlop} et~al.}{2003}]{2003dunlop}
{Dunlop} J.~S.,  {McLure} R.~J.,  {Kukula} M.~J.,  {Baum} S.~A.,  {O'Dea}
  C.~P.,   {Hughes} D.~H.,  2003, \mn@doi [\mnras]
  {10.1046/j.1365-8711.2003.06333.x}, \href
  {https://ui.adsabs.harvard.edu/abs/2003MNRAS.340.1095D} {340, 1095}

\bibitem[\protect\citeauthoryear{{Dyba}, {Mach}  \& {Malec}}{{Dyba}
  et~al.}{2019}]{2019dyba}
{Dyba} W.,  {Mach} P.,   {Malec} E.,  2019, \mn@doi [\mnras]
  {10.1093/mnras/stz1058}, \href
  {https://ui.adsabs.harvard.edu/abs/2019MNRAS.486.3118D} {486, 3118}

\bibitem[\protect\citeauthoryear{{Einstein}, {Infeld}  \&
  {Hoffmann}}{{Einstein} et~al.}{1938}]{1938einstein}
{Einstein} A.,  {Infeld} L.,   {Hoffmann} B.,  1938, Annals of Mathematics,
  \href {https://ui.adsabs.harvard.edu/abs/1938AnMat..39...65E} {39, 65}

\bibitem[\protect\citeauthoryear{{Fan}, {Liu}, {Qian}, {Tao}, {Shen}, {Zhang},
  {Huang}  \& {Wang}}{{Fan} et~al.}{2010}]{2010fan}
{Fan} J.-H.,  {Liu} Y.,  {Qian} B.-C.,  {Tao} J.,  {Shen} Z.-Q.,  {Zhang}
  J.-S.,  {Huang} Y.,   {Wang} J.,  2010, \mn@doi [Research in Astronomy and
  Astrophysics] {10.1088/1674-4527/10/11/002}, \href
  {https://ui.adsabs.harvard.edu/abs/2010RAA....10.1100F} {10, 1100}

\bibitem[\protect\citeauthoryear{{Faye}, {Blanchet}  \& {Buonanno}}{{Faye}
  et~al.}{2006}]{2006faye}
{Faye} G.,  {Blanchet} L.,   {Buonanno} A.,  2006, \mn@doi [\prd]
  {10.1103/PhysRevD.74.104033}, \href
  {https://ui.adsabs.harvard.edu/abs/2006PhRvD..74j4033F} {74, 104033}

\bibitem[\protect\citeauthoryear{Foreman-Mackey}{Foreman-Mackey}{2016}]{corner}
Foreman-Mackey D.,  2016, \mn@doi [The Journal of Open Source Software]
  {10.21105/joss.00024}, 1, 24

\bibitem[\protect\citeauthoryear{{Foreman-Mackey}, {Hogg}, {Lang}  \&
  {Goodman}}{{Foreman-Mackey} et~al.}{2013}]{2013emcee}
{Foreman-Mackey} D.,  {Hogg} D.~W.,  {Lang} D.,   {Goodman} J.,  2013, \mn@doi
  [\pasp] {10.1086/670067}, \href
  {https://ui.adsabs.harvard.edu/abs/2013PASP..125..306F} {125, 306}

\bibitem[\protect\citeauthoryear{{Franchini} et~al.,}{{Franchini}
  et~al.}{2023}]{2023franchini}
{Franchini} A.,  et~al., 2023, \mn@doi [\aap] {10.1051/0004-6361/202346565},
  \href {https://ui.adsabs.harvard.edu/abs/2023A&A...675A.100F} {675, A100}

\bibitem[\protect\citeauthoryear{{Gaida} \& {Roeser}}{{Gaida} \&
  {Roeser}}{1982}]{1982gaida}
{Gaida} G.,  {Roeser} H.~J.,  1982, \aap, \href
  {https://ui.adsabs.harvard.edu/abs/1982A&A...105..362G} {105, 362}

\bibitem[\protect\citeauthoryear{{Gelman} \& {Rubin}}{{Gelman} \&
  {Rubin}}{1992}]{1992gelman}
{Gelman} A.,  {Rubin} D.~B.,  1992, \mn@doi [Statistical Science]
  {10.1214/ss/1177011136}, \href
  {https://ui.adsabs.harvard.edu/abs/1992StaSc...7..457G} {7, 457}

\bibitem[\protect\citeauthoryear{{Ghisellini}, {Celotti}, {Fossati}, {Maraschi}
   \& {Comastri}}{{Ghisellini} et~al.}{1998}]{1998Ghisellini}
{Ghisellini} G.,  {Celotti} A.,  {Fossati} G.,  {Maraschi} L.,   {Comastri} A.,
   1998, \mn@doi [\mnras] {10.1046/j.1365-8711.1998.02032.x}, \href
  {https://ui.adsabs.harvard.edu/abs/1998MNRAS.301..451G} {301, 451}

\bibitem[\protect\citeauthoryear{{Gierli{\'n}ski} \& {Done}}{{Gierli{\'n}ski}
  \& {Done}}{2004}]{2004gierli}
{Gierli{\'n}ski} M.,  {Done} C.,  2004, \mn@doi [\mnras]
  {10.1111/j.1365-2966.2004.07266.x}, \href
  {https://ui.adsabs.harvard.edu/abs/2004MNRAS.347..885G} {347, 885}

\bibitem[\protect\citeauthoryear{{Gupta} et~al.,}{{Gupta}
  et~al.}{2017}]{2017gupta}
{Gupta} A.~C.,  et~al., 2017, \mn@doi [\mnras] {10.1093/mnras/stw3045}, \href
  {https://ui.adsabs.harvard.edu/abs/2017MNRAS.465.4423G} {465, 4423}

\bibitem[\protect\citeauthoryear{{Hudec}, {Ba{\v{s}}ta}, {Pihajoki}  \&
  {Valtonen}}{{Hudec} et~al.}{2013}]{2013Hudec}
{Hudec} R.,  {Ba{\v{s}}ta} M.,  {Pihajoki} P.,   {Valtonen} M.,  2013, \mn@doi
  [\aap] {10.1051/0004-6361/201219323}, \href
  {https://ui.adsabs.harvard.edu/abs/2013A&A...559A..20H} {559, A20}

\bibitem[\protect\citeauthoryear{{Iorio} \& {Zhang}}{{Iorio} \&
  {Zhang}}{2017}]{2017iorio}
{Iorio} L.,  {Zhang} F.,  2017, \mn@doi [\apj] {10.3847/1538-4357/aa671b},
  \href {https://ui.adsabs.harvard.edu/abs/2017ApJ...839....3I} {839, 3}

\bibitem[\protect\citeauthoryear{{Ivanov}, {Igumenshchev}  \&
  {Novikov}}{{Ivanov} et~al.}{1998}]{1998ivanov}
{Ivanov} P.~B.,  {Igumenshchev} I.~V.,   {Novikov} I.~D.,  1998, \mn@doi [\apj]
  {10.1086/306324}, \href
  {https://ui.adsabs.harvard.edu/abs/1998ApJ...507..131I} {507, 131}

\bibitem[\protect\citeauthoryear{{Jiang} \& {Blaes}}{{Jiang} \&
  {Blaes}}{2020}]{2020jiang}
{Jiang} Y.-F.,  {Blaes} O.,  2020, \mn@doi [\apj] {10.3847/1538-4357/aba4b7},
  \href {https://ui.adsabs.harvard.edu/abs/2020ApJ...900...25J} {900, 25}

\bibitem[\protect\citeauthoryear{{Jiang}, {Davis}  \& {Stone}}{{Jiang}
  et~al.}{2016}]{2016Jiang}
{Jiang} Y.-F.,  {Davis} S.~W.,   {Stone} J.~M.,  2016, \mn@doi [\apj]
  {10.3847/0004-637X/827/1/10}, \href
  {https://ui.adsabs.harvard.edu/abs/2016ApJ...827...10J} {827, 10}

\bibitem[\protect\citeauthoryear{{Kacskovics} \& {Vas{\'u}th}}{{Kacskovics} \&
  {Vas{\'u}th}}{2022}]{2022kacskovics}
{Kacskovics} B.,  {Vas{\'u}th} M.,  2022, \mn@doi [Classical and Quantum
  Gravity] {10.1088/1361-6382/ac5d17}, \href
  {https://ui.adsabs.harvard.edu/abs/2022CQGra..39i5007K} {39, 095007}

\bibitem[\protect\citeauthoryear{{Karas} \& {Vokrouhlicky}}{{Karas} \&
  {Vokrouhlicky}}{1994}]{1994karas}
{Karas} V.,  {Vokrouhlicky} D.,  1994, \mn@doi [\apj] {10.1086/173719}, \href
  {https://ui.adsabs.harvard.edu/abs/1994ApJ...422..208K} {422, 208}

\bibitem[\protect\citeauthoryear{{Katz}}{{Katz}}{1997}]{1997katz}
{Katz} J.~I.,  1997, \mn@doi [\apj] {10.1086/303811}, \href
  {https://ui.adsabs.harvard.edu/abs/1997ApJ...478..527K} {478, 527}

\bibitem[\protect\citeauthoryear{{Kidder}}{{Kidder}}{1995}]{1995kidder}
{Kidder} L.~E.,  1995, \mn@doi [\prd] {10.1103/PhysRevD.52.821}, \href
  {https://ui.adsabs.harvard.edu/abs/1995PhRvD..52..821K} {52, 821}

\bibitem[\protect\citeauthoryear{{King}, {Lubow}, {Ogilvie}  \&
  {Pringle}}{{King} et~al.}{2005}]{2005king}
{King} A.~R.,  {Lubow} S.~H.,  {Ogilvie} G.~I.,   {Pringle} J.~E.,  2005,
  \mn@doi [\mnras] {10.1111/j.1365-2966.2005.09378.x}, \href
  {https://ui.adsabs.harvard.edu/abs/2005MNRAS.363...49K} {363, 49}

\bibitem[\protect\citeauthoryear{{King}, {Pringle}  \& {Livio}}{{King}
  et~al.}{2007}]{2007king}
{King} A.~R.,  {Pringle} J.~E.,   {Livio} M.,  2007, \mn@doi [\mnras]
  {10.1111/j.1365-2966.2007.11556.x}, \href
  {https://ui.adsabs.harvard.edu/abs/2007MNRAS.376.1740K} {376, 1740}

\bibitem[\protect\citeauthoryear{{Kinman} \& {Conklin}}{{Kinman} \&
  {Conklin}}{1971}]{1971Kinman}
{Kinman} T.~D.,  {Conklin} E.~K.,  1971, \aplett, \href
  {https://ui.adsabs.harvard.edu/abs/1971ApL.....9..147K} {9, 147}

\bibitem[\protect\citeauthoryear{{Komossa} et~al.,}{{Komossa}
  et~al.}{2015}]{2015komossa}
{Komossa} S.,  et~al., 2015, The Astronomer's Telegram, \href
  {https://ui.adsabs.harvard.edu/abs/2015ATel.8411....1K} {8411, 1}

\bibitem[\protect\citeauthoryear{{Komossa} et~al.,}{{Komossa}
  et~al.}{2023a}]{2023komossa}
{Komossa} S.,  et~al., 2023a, \mn@doi [\mnras] {10.1093/mnrasl/slad016}, \href
  {https://ui.adsabs.harvard.edu/abs/2023MNRAS.tmpL..21K} {}

\bibitem[\protect\citeauthoryear{{Komossa} et~al.,}{{Komossa}
  et~al.}{2023b}]{2023komossa2}
{Komossa} S.,  et~al., 2023b, \mn@doi [\apj] {10.3847/1538-4357/acaf71}, \href
  {https://ui.adsabs.harvard.edu/abs/2023ApJ...944..177K} {944, 177}

\bibitem[\protect\citeauthoryear{{Kotko} \& {Lasota}}{{Kotko} \&
  {Lasota}}{2012}]{2012kotko}
{Kotko} I.,  {Lasota} J.~P.,  2012, \mn@doi [\aap]
  {10.1051/0004-6361/201219618}, \href
  {https://ui.adsabs.harvard.edu/abs/2012A&A...545A.115K} {545, A115}

\bibitem[\protect\citeauthoryear{{Kuzmin} \& {Malasidze}}{{Kuzmin} \&
  {Malasidze}}{1987}]{1987Kuzmin}
{Kuzmin} G.~G.,  {Malasidze} G.~A.,  1987, Publications of the Tartu
  Astrofizica Observatory, \href
  {https://ui.adsabs.harvard.edu/abs/1987PTarO..52...48K} {52, 48}

\bibitem[\protect\citeauthoryear{{Laine} et~al.,}{{Laine}
  et~al.}{2020}]{2020Laine}
{Laine} S.,  et~al., 2020, \mn@doi [\apjl] {10.3847/2041-8213/ab79a4}, \href
  {https://ui.adsabs.harvard.edu/abs/2020ApJ...894L...1L} {894, L1}

\bibitem[\protect\citeauthoryear{{Lehto} \& {Valtonen}}{{Lehto} \&
  {Valtonen}}{1996}]{LV96}
{Lehto} H.~J.,  {Valtonen} M.~J.,  1996, \mn@doi [\apj] {10.1086/176962}, \href
  {https://ui.adsabs.harvard.edu/abs/1996ApJ...460..207L} {460, 207}

\bibitem[\protect\citeauthoryear{{Lema{\^\i}tre}}{{Lema{\^\i}tre}}{1931}]{1931lemaitre}
{Lema{\^\i}tre} G.,  1931, \mn@doi [\mnras] {10.1093/mnras/91.5.483}, \href
  {https://ui.adsabs.harvard.edu/abs/1931MNRAS..91..483L} {91, 483}

\bibitem[\protect\citeauthoryear{{Maggiore}}{{Maggiore}}{2018}]{2018maggiore}
{Maggiore} M.,  2018, {Gravitational Waves: Volume 2: Astrophysics and
  Cosmology}, \mn@doi{10.1093/oso/9780198570899.001.0001.
}

\bibitem[\protect\citeauthoryear{{McKerns}, {Strand}, {Sullivan}, {Fang}  \&
  {Aivazis}}{{McKerns} et~al.}{2012}]{2012multi}
{McKerns} M.~M.,  {Strand} L.,  {Sullivan} T.,  {Fang} A.,   {Aivazis} M.
  A.~G.,  2012, \mn@doi [arXiv e-prints] {10.48550/arXiv.1202.1056}, \href
  {https://ui.adsabs.harvard.edu/abs/2012arXiv1202.1056M} {p. arXiv:1202.1056}

\bibitem[\protect\citeauthoryear{{Natarajan} \& {Pringle}}{{Natarajan} \&
  {Pringle}}{1998}]{1998natarajan}
{Natarajan} P.,  {Pringle} J.~E.,  1998, \mn@doi [\apjl] {10.1086/311658},
  \href {https://ui.adsabs.harvard.edu/abs/1998ApJ...506L..97N} {506, L97}

\bibitem[\protect\citeauthoryear{{Peters} \& {Mathews}}{{Peters} \&
  {Mathews}}{1963}]{1963peters}
{Peters} P.~C.,  {Mathews} J.,  1963, \mn@doi [Physical Review]
  {10.1103/PhysRev.131.435}, \href
  {https://ui.adsabs.harvard.edu/abs/1963PhRv..131..435P} {131, 435}

\bibitem[\protect\citeauthoryear{{Pietil{\"a}}}{{Pietil{\"a}}}{1998}]{1998Pietila}
{Pietil{\"a}} H.,  1998, \mn@doi [\apj] {10.1086/306444}, \href
  {https://ui.adsabs.harvard.edu/abs/1998ApJ...508..669P} {508, 669}

\bibitem[\protect\citeauthoryear{{Pihajoki}}{{Pihajoki}}{2016}]{2016pihajoki}
{Pihajoki} P.,  2016, \mn@doi [\mnras] {10.1093/mnras/stv3023}, \href
  {https://ui.adsabs.harvard.edu/abs/2016MNRAS.457.1145P} {457, 1145}

\bibitem[\protect\citeauthoryear{{Pihajoki}, {Valtonen}  \&
  {Ciprini}}{{Pihajoki} et~al.}{2013a}]{2013pihajoki}
{Pihajoki} P.,  {Valtonen} M.,   {Ciprini} S.,  2013a, \mn@doi [\mnras]
  {10.1093/mnras/stt1233}, \href
  {https://ui.adsabs.harvard.edu/abs/2013MNRAS.434.3122P} {434, 3122}

\bibitem[\protect\citeauthoryear{{Pihajoki} et~al.,}{{Pihajoki}
  et~al.}{2013b}]{2013pihajoki2}
{Pihajoki} P.,  et~al., 2013b, \mn@doi [\apj] {10.1088/0004-637X/764/1/5},
  \href {https://ui.adsabs.harvard.edu/abs/2013ApJ...764....5P} {764, 5}

\bibitem[\protect\citeauthoryear{{Porto}}{{Porto}}{2006}]{2006porto}
{Porto} R.~A.,  2006, \mn@doi [\prd] {10.1103/PhysRevD.73.104031}, \href
  {https://ui.adsabs.harvard.edu/abs/2006PhRvD..73j4031P} {73, 104031}

\bibitem[\protect\citeauthoryear{{Qian}}{{Qian}}{2018}]{2018qian}
{Qian} S.,  2018, \mn@doi [arXiv e-prints] {10.48550/arXiv.1811.11514}, \href
  {https://ui.adsabs.harvard.edu/abs/2018arXiv181111514Q} {p. arXiv:1811.11514}

\bibitem[\protect\citeauthoryear{{Sakimoto} \& {Coroniti}}{{Sakimoto} \&
  {Coroniti}}{1981}]{1981sakimoto}
{Sakimoto} P.~J.,  {Coroniti} F.~V.,  1981, \mn@doi [\apj] {10.1086/159005},
  \href {https://ui.adsabs.harvard.edu/abs/1981ApJ...247...19S} {247, 19}

\bibitem[\protect\citeauthoryear{{Sch{\"a}fer} \& {Jaranowski}}{{Sch{\"a}fer}
  \& {Jaranowski}}{2018}]{2018schaefer}
{Sch{\"a}fer} G.,  {Jaranowski} P.,  2018, \mn@doi [Living Reviews in
  Relativity] {10.1007/s41114-018-0016-5}, \href
  {https://ui.adsabs.harvard.edu/abs/2018LRR....21....7S} {21, 7}

\bibitem[\protect\citeauthoryear{{Shakura} \& {Sunyaev}}{{Shakura} \&
  {Sunyaev}}{1973}]{1973shakura}
{Shakura} N.~I.,  {Sunyaev} R.~A.,  1973, \aap, \href
  {https://ui.adsabs.harvard.edu/abs/1973A&A....24..337S} {24, 337}

\bibitem[\protect\citeauthoryear{{Shapiro}}{{Shapiro}}{1964}]{1964shapiro}
{Shapiro} I.~I.,  1964, \mn@doi [\prl] {10.1103/PhysRevLett.13.789}, \href
  {https://ui.adsabs.harvard.edu/abs/1964PhRvL..13..789S} {13, 789}

\bibitem[\protect\citeauthoryear{{Shappee} et~al.,}{{Shappee}
  et~al.}{2015}]{2015shappe}
{Shappee} B.~J.,  et~al., 2015, The Astronomer's Telegram, \href
  {https://ui.adsabs.harvard.edu/abs/2015ATel.8372....1S} {8372, 1}

\bibitem[\protect\citeauthoryear{{Sillanpaa}, {Haarala}, {Valtonen},
  {Sundelius}  \& {Byrd}}{{Sillanpaa} et~al.}{1988}]{1988Sillanpää}
{Sillanpaa} A.,  {Haarala} S.,  {Valtonen} M.~J.,  {Sundelius} B.,   {Byrd}
  G.~G.,  1988, \mn@doi [\apj] {10.1086/166033}, \href
  {https://ui.adsabs.harvard.edu/abs/1988ApJ...325..628S} {325, 628}

\bibitem[\protect\citeauthoryear{{Sitko} \& {Junkkarinen}}{{Sitko} \&
  {Junkkarinen}}{1985}]{1985sitko}
{Sitko} M.~L.,  {Junkkarinen} V.~T.,  1985, \mn@doi [\pasp] {10.1086/131679},
  \href {https://ui.adsabs.harvard.edu/abs/1985PASP...97.1158S} {97, 1158}

\bibitem[\protect\citeauthoryear{{Stella} \& {Rosner}}{{Stella} \&
  {Rosner}}{1984}]{1984rosner}
{Stella} L.,  {Rosner} R.,  1984, \mn@doi [\apj] {10.1086/161697}, \href
  {https://ui.adsabs.harvard.edu/abs/1984ApJ...277..312S} {277, 312}

\bibitem[\protect\citeauthoryear{{Sundelius}, {Wahde}, {Lehto}  \&
  {Valtonen}}{{Sundelius} et~al.}{1997}]{1997Sundelius}
{Sundelius} B.,  {Wahde} M.,  {Lehto} H.~J.,   {Valtonen} M.~J.,  1997, \mn@doi
  [\apj] {10.1086/304331}, \href
  {https://ui.adsabs.harvard.edu/abs/1997ApJ...484..180S} {484, 180}

\bibitem[\protect\citeauthoryear{{Tang}, {Zhang}  \& {Pang}}{{Tang}
  et~al.}{2014}]{2014JTang}
{Tang} J.,  {Zhang} H.-J.,   {Pang} Q.,  2014, \mn@doi [Journal of Astrophysics
  and Astronomy] {10.1007/s12036-014-9218-8}, \href
  {https://ui.adsabs.harvard.edu/abs/2014JApA...35..301T} {35, 301}

\bibitem[\protect\citeauthoryear{{Titarchuk}, {Seifina}  \&
  {Shrader}}{{Titarchuk} et~al.}{2023}]{2023titarchuk}
{Titarchuk} L.,  {Seifina} E.,   {Shrader} C.,  2023, \mn@doi [\aap]
  {10.1051/0004-6361/202345923}, \href
  {https://ui.adsabs.harvard.edu/abs/2023A&A...671A.159T} {671, A159}

\bibitem[\protect\citeauthoryear{{Toomre}}{{Toomre}}{1964}]{1964toomre}
{Toomre} A.,  1964, \mn@doi [\apj] {10.1086/147861}, \href
  {https://ui.adsabs.harvard.edu/abs/1964ApJ...139.1217T} {139, 1217}

\bibitem[\protect\citeauthoryear{{Urry} \& {Padovani}}{{Urry} \&
  {Padovani}}{1995}]{1995Urry}
{Urry} C.~M.,  {Padovani} P.,  1995, \mn@doi [\pasp] {10.1086/133630}, \href
  {https://ui.adsabs.harvard.edu/abs/1995PASP..107..803U} {107, 803}

\bibitem[\protect\citeauthoryear{{Valtaoja} et~al.,}{{Valtaoja}
  et~al.}{1985}]{1985valtaoja}
{Valtaoja} E.,  et~al., 1985, \mn@doi [\nat] {10.1038/314148a0}, \href
  {https://ui.adsabs.harvard.edu/abs/1985Natur.314..148V} {314, 148}

\bibitem[\protect\citeauthoryear{{Valtaoja}, {Ter{\"a}sranta}, {Tornikoski},
  {Sillanp{\"a}{\"a}}, {Aller}, {Aller}  \& {Hughes}}{{Valtaoja}
  et~al.}{2000}]{2000Valtaoja}
{Valtaoja} E.,  {Ter{\"a}sranta} H.,  {Tornikoski} M.,  {Sillanp{\"a}{\"a}} A.,
   {Aller} M.~F.,  {Aller} H.~D.,   {Hughes} P.~A.,  2000, \mn@doi [\apj]
  {10.1086/308494}, \href
  {https://ui.adsabs.harvard.edu/abs/2000ApJ...531..744V} {531, 744}

\bibitem[\protect\citeauthoryear{{Valtonen}}{{Valtonen}}{2007}]{2007valtonen}
{Valtonen} M.~J.,  2007, \mn@doi [\apj] {10.1086/512801}, \href
  {https://ui.adsabs.harvard.edu/abs/2007ApJ...659.1074V} {659, 1074}

\bibitem[\protect\citeauthoryear{{Valtonen} et~al.,}{{Valtonen}
  et~al.}{2006a}]{2006valtonen2}
{Valtonen} M.~J.,  et~al., 2006a, \mn@doi [\apjl] {10.1086/505039}, \href
  {https://ui.adsabs.harvard.edu/abs/2006ApJ...643L...9V} {643, L9}

\bibitem[\protect\citeauthoryear{{Valtonen} et~al.,}{{Valtonen}
  et~al.}{2006b}]{2006valtonen}
{Valtonen} M.~J.,  et~al., 2006b, \mn@doi [\apj] {10.1086/504884}, \href
  {https://ui.adsabs.harvard.edu/abs/2006ApJ...646...36V} {646, 36}

\bibitem[\protect\citeauthoryear{{Valtonen} et~al.,}{{Valtonen}
  et~al.}{2010}]{2010valtonen}
{Valtonen} M.~J.,  et~al., 2010, \mn@doi [\apj] {10.1088/0004-637X/709/2/725},
  \href {https://ui.adsabs.harvard.edu/abs/2010ApJ...709..725V} {709, 725}

\bibitem[\protect\citeauthoryear{{Valtonen} et~al.,}{{Valtonen}
  et~al.}{2016}]{2016valtonen}
{Valtonen} M.~J.,  et~al., 2016, \mn@doi [\apjl] {10.3847/2041-8205/819/2/L37},
  \href {https://ui.adsabs.harvard.edu/abs/2016ApJ...819L..37V} {819, L37}

\bibitem[\protect\citeauthoryear{{Valtonen} et~al.,}{{Valtonen}
  et~al.}{2019}]{2019valtonendisc}
{Valtonen} M.~J.,  et~al., 2019, \mn@doi [\apj] {10.3847/1538-4357/ab3573},
  \href {https://ui.adsabs.harvard.edu/abs/2019ApJ...882...88V} {882, 88}

\bibitem[\protect\citeauthoryear{{Valtonen} et~al.,}{{Valtonen}
  et~al.}{2021}]{2021Valtonen}
{Valtonen} M.~J.,  et~al., 2021, \mn@doi [Galaxies] {10.3390/galaxies10010001},
  \href {https://ui.adsabs.harvard.edu/abs/2021Galax..10....1V} {10, 1}

\bibitem[\protect\citeauthoryear{{Valtonen} et~al.,}{{Valtonen}
  et~al.}{2022}]{2022valtonen}
{Valtonen} M.~J.,  et~al., 2022, \mn@doi [arXiv e-prints]
  {10.48550/arXiv.2209.08360}, \href
  {https://ui.adsabs.harvard.edu/abs/2022arXiv220908360V} {p. arXiv:2209.08360}

\bibitem[\protect\citeauthoryear{{Valtonen} et~al.,}{{Valtonen}
  et~al.}{2023a}]{2023valtonendiss}
{Valtonen} M.~J.,  et~al., 2023a, \mn@doi [Galaxies]
  {10.3390/galaxies11040082}, \href
  {https://ui.adsabs.harvard.edu/abs/2023Galax..11...82V} {11, 82}

\bibitem[\protect\citeauthoryear{{Valtonen} et~al.,}{{Valtonen}
  et~al.}{2023b}]{2023valtonen}
{Valtonen} M.~J.,  et~al., 2023b, \mn@doi [\mnras] {10.1093/mnras/stad922},
  \href {https://ui.adsabs.harvard.edu/abs/2023MNRAS.521.6143V} {521, 6143}

\bibitem[\protect\citeauthoryear{{Vats} \& {Knudson}}{{Vats} \&
  {Knudson}}{2018}]{2018vats}
{Vats} D.,  {Knudson} C.,  2018, \mn@doi [arXiv e-prints]
  {10.48550/arXiv.1812.09384}, \href
  {https://ui.adsabs.harvard.edu/abs/2018arXiv181209384V} {p. arXiv:1812.09384}

\bibitem[\protect\citeauthoryear{{Villata}, {Raiteri}, {Sillanpaa}  \&
  {Takalo}}{{Villata} et~al.}{1998}]{1998Villata}
{Villata} M.,  {Raiteri} C.~M.,  {Sillanpaa} A.,   {Takalo} L.~O.,  1998,
  \mn@doi [\mnras] {10.1046/j.1365-8711.1998.01244.x}, \href
  {https://ui.adsabs.harvard.edu/abs/1998MNRAS.293L..13V} {293, L13}

\bibitem[\protect\citeauthoryear{{Villforth} et~al.,}{{Villforth}
  et~al.}{2010}]{2010Villforth}
{Villforth} C.,  et~al., 2010, \mn@doi [\mnras]
  {10.1111/j.1365-2966.2009.16133.x}, \href
  {https://ui.adsabs.harvard.edu/abs/2010MNRAS.402.2087V} {402, 2087}

\bibitem[\protect\citeauthoryear{Virtanen et~al.,}{Virtanen
  et~al.}{2020}]{2020scipy}
Virtanen P.,  et~al., 2020, \mn@doi [Nature Methods]
  {10.1038/s41592-019-0686-2}, \href {https://rdcu.be/b08Wh} {17, 261}

\bibitem[\protect\citeauthoryear{{Volonteri}, {Madau}, {Quataert}  \&
  {Rees}}{{Volonteri} et~al.}{2005}]{2005volonteri}
{Volonteri} M.,  {Madau} P.,  {Quataert} E.,   {Rees} M.~J.,  2005, \mn@doi
  [\apj] {10.1086/426858}, \href
  {https://ui.adsabs.harvard.edu/abs/2005ApJ...620...69V} {620, 69}

\bibitem[\protect\citeauthoryear{{Will}}{{Will}}{2014a}]{2014will2}
{Will} C.~M.,  2014a, \mn@doi [Classical and Quantum Gravity]
  {10.1088/0264-9381/31/24/244001}, \href
  {https://ui.adsabs.harvard.edu/abs/2014CQGra..31x4001W} {31, 244001}

\bibitem[\protect\citeauthoryear{{Will}}{{Will}}{2014b}]{2014will}
{Will} C.~M.,  2014b, \mn@doi [\prd] {10.1103/PhysRevD.89.044043}, \href
  {https://ui.adsabs.harvard.edu/abs/2014PhRvD..89d4043W} {89, 044043}

\bibitem[\protect\citeauthoryear{{Will} \& {Maitra}}{{Will} \&
  {Maitra}}{2017}]{2017will}
{Will} C.~M.,  {Maitra} M.,  2017, \mn@doi [\prd] {10.1103/PhysRevD.95.064003},
  \href {https://ui.adsabs.harvard.edu/abs/2017PhRvD..95f4003W} {95, 064003}

\bibitem[\protect\citeauthoryear{{Wolf}}{{Wolf}}{1916}]{1916Wölf}
{Wolf} M.,  1916, Astronomische Nachrichten, \href
  {https://ui.adsabs.harvard.edu/abs/1916AN....202..415W} {202, 415}

\bibitem[\protect\citeauthoryear{{Yanny}, {Jannuzi}  \& {Impey}}{{Yanny}
  et~al.}{1997}]{1997Yanny}
{Yanny} B.,  {Jannuzi} B.~T.,   {Impey} C.,  1997, \mn@doi [\apjl]
  {10.1086/310793}, \href
  {https://ui.adsabs.harvard.edu/abs/1997ApJ...484L.113Y} {484, L113}

\bibitem[\protect\citeauthoryear{{Zanotti}, {Roedig}, {Rezzolla}  \& {Del
  Zanna}}{{Zanotti} et~al.}{2011}]{2011Zanotti}
{Zanotti} O.,  {Roedig} C.,  {Rezzolla} L.,   {Del Zanna} L.,  2011, \mn@doi
  [\mnras] {10.1111/j.1365-2966.2011.19451.x}, \href
  {https://ui.adsabs.harvard.edu/abs/2011MNRAS.417.2899Z} {417, 2899}

\bibitem[\protect\citeauthoryear{{Zwick}, {Capelo}, {Bortolas},
  {V{\'a}zquez-Aceves}, {Mayer}  \& {Amaro-Seoane}}{{Zwick}
  et~al.}{2021}]{2021zwick}
{Zwick} L.,  {Capelo} P.~R.,  {Bortolas} E.,  {V{\'a}zquez-Aceves} V.,  {Mayer}
  L.,   {Amaro-Seoane} P.,  2021, \mn@doi [\mnras] {10.1093/mnras/stab1818},
  \href {https://ui.adsabs.harvard.edu/abs/2021MNRAS.506.1007Z} {506, 1007}

\bibitem[\protect\citeauthoryear{{de Diego} \& {Kidger}}{{de Diego} \&
  {Kidger}}{1990}]{1990dediego}
{de Diego} J.~A.,  {Kidger} M.,  1990, \mn@doi [\apss] {10.1007/BF00646827},
  \href {https://ui.adsabs.harvard.edu/abs/1990Ap&SS.171...97D} {171, 97}

\makeatother
\end{thebibliography}
\normalsize

\bsp 
\label{lastpage}
\end{document}